\documentclass[12pt,a4paper]{article} 
\usepackage{jheppub}

\newcommand{\bea}{\begin{eqnarray}}
\newcommand{\eea}{\end{eqnarray}}

\newcommand{\beq}{\begin{equation}}
\newcommand{\eeq}{\end{equation}}
\newcommand{\beqa}{\begin{eqnarray}}
\newcommand{\eeqa}{\end{eqnarray}}

\def\x{{\boldsymbol x}}

\def\k{{\boldsymbol k}}

\def\q{{\boldsymbol q}}
\def\Q{{\boldsymbol Q}}

\def\0{{\boldsymbol 0}}

\def\K{{\boldsymbol K}}
\def\J{{\boldsymbol J}}

\def\cal{\mathcal}

\def\kg{k_g}
\def\xg{x_g}

\def\kkg{{\boldsymbol k}_g}

\def\eg{\epsilon_g}
\def\ei{\epsilon_i}
\def\ef{\epsilon_f}

\def\eeg{{\boldsymbol \epsilon}_g}
\def\eei{{\boldsymbol \epsilon}_i}
\def\eef{{\boldsymbol \epsilon}_f}
\def\wob{\overline{\omega}_0}
\def\wib{\overline{\omega}_1}
\def\wiib{\overline{\omega}_2}
\def\woiib{\overline{\omega}_{02}}

\relax

\title{The contribution of medium-modified color flow to jet quenching}

\author[a]{Andrea Beraudo}

\author[a,b]{Jos\'e Guilherme Milhano}

\author[a]{Urs Achim Wiedemann}

\affiliation[a]{Physics Department, Theory Unit, CERN, CH-1211 Gen\`eve 23, Switzerland}

\affiliation[b]{CENTRA, Instituto Superior T\'ecnico, Universidade T\'ecnica de Lisboa, \\ Av. Rovisco Pais 1, P-1049-001 Lisboa, Portugal}

\emailAdd{andrea.beraudo@cern.ch}
\emailAdd{guilherme.milhano@ist.utl.pt}
\emailAdd{urs.wiedemann@cern.ch}

\abstract{Multiple interactions between parton showers and the surrounding QCD matter are expected to underlie the strong 
medium-modifications of jet observables in ultra-relativistic heavy ion collisions at RHIC and at the LHC. Here, we note that such jet-medium 
interactions alter generically and characteristically the color correlations in the parton shower. We characterize these effects in a color-differential 
calculation of the medium-induced gluon radiation spectrum to first and second order in opacity. By interfacing simple branching histories of 
medium-modified color flow with the \textsc{Lund} hadronization model, we analyze how the medium modification of color correlations can affect the 
distribution of hadronic fragments in jets. Importantly, we observe that jet-medium interactions give rise to the medium-induced color decoherence
of gluons from the parton shower. Since hadronization respects color flow and since each color singlet in a parton shower is hadronized separately, 
this medium-induced color decoherence leaves characteristic signatures in the jet fragmentation pattern. In particular, it can contribute to the
quenching of leading hadron spectra. Moreover, it can
increase strongly the yield of soft hadronic fragments from a jet, while the distribution of more energetic hadrons follows naturally the
shape of a vacuum-like fragmentation pattern of lower total energy. 
}

\preprint{CERN-PH-TH/2012-100}

\begin{document}


\maketitle

\section{Introduction}
In ultra-relativistic nucleus-nucleus collisions at RHIC and at the LHC essentially all hadronic particle distributions at high transverse momentum 
show strong modifications if compared to baselines established in proton-proton collisions. In particular, one observes a strong suppression of all single
inclusive hadron spectra up to the highest transverse momenta ($p_T \sim 100$ GeV) analyzed so far \cite{Aamodt:2010jd,CMS:2012aa}, and significant modifications of jet structures
up to jet energies exceeding 300 GeV~\cite{Aad:2010bu,Chatrchyan:2011sx,Chatrchyan:2012ni}. Characteristic features of this jet quenching phenomenon include its dependence on centrality and azimuthal orientation (both of which yielding information on the path-length of  in-medium propagation), its dependence on kinematic variables including the center of mass energy of the collision and the transverse momentum of the hard process, its approximate independence on the produced hadron species (at least in the limited low-$p_T$ range in which the latter has been measured to date), and the absence of quenching effect for high-$p_T$ prompt photons and $Z$-bosons. The totality of these data from LHC  and the data from RHIC~\cite{Adams:2005dq,Adcox:2004mh} motivates a dynamical picture of jet quenching according to which partons 
are produced in nucleus-nucleus collisions via high-momentum transfer processes at standard perturbative rates, but lose energy and branch differently in the dense QCD medium through which they propagate. Starting with the seminal works of Baier, Dokshitzer, Mueller, Peign\'e and Schiff (BDMPS) \cite{Baier:1996sk}and Zakharov (Z) \cite{Zakharov:1997uu} in the 1990s, a large number of parton energy loss calculations aim at formulating and exploring this jet quenching phenomenon in a QCD-based setup (for recent reviews, see \cite{CasalderreySolana:2007zz,Wiedemann:2009sh,Majumder:2010qh,Armesto:2011ht}) 

In parton energy loss calculations, one generally considers a class of processes in which high-energy partons, produced in nucleus-nucleus collisions, interact via gluon exchanges with the surrounding QCD matter while branching. Color exchange between the partonic projectile and the QCD medium is intrinsic to such interactions. As a consequence, the color connection within a high-energy parton shower and between the shower and the rest of the event will be modified by the medium. Hadronization respects color correlations in mapping partonic color-singlet configurations into hadrons and can thus be sensitive to medium-modified color flow.  Since the input of any hadronization routine is different if the color connections are different, medium modifications of the hadronization process may be expected to persist even if the latter occurs time delayed and thus outside the QCD medium. 
Taking such qualitative considerations into account, several groups have explored heuristic models of medium-modified hadronization in the recent past, pointing to possible changes in the hadrochemical composition of jets~\cite{Sapeta:2007ad} or in the yield and distribution of baryons~\cite{Aurenche:2011rd}. 
Also, some basic scenarios of medium-modified hadronization were explored in Monte Carlo simulations of medium-induced parton energy loss~\cite{Zapp:2008gi}. Notwithstanding these efforts, QCD-based calculations of jet quenching have remained mainly focused on kinematic changes of parton branching. Here, we extend the calculation of medium-induced gluon radiation of BDMPS-Z to the study of the color-differential case.
We investigate, in particular,  which medium-modified color connections can arise in the interactions of a parton shower with a QCD medium. Based on these calculations, we shall then test the response of hadronization models to a medium-modified color flow and discuss conceivable experimental signatures. 

Our work is organized as follows. In section~\ref{sec2}, we set the stage by discussing qualitative features of the color flow of a parton shower developing in a QCD medium and how medium-modified color connections would affect the input of standard hadronization models. Sections~\ref{sec3} and ~\ref{sec4} provide an explicit color-differential analysis of the medium-induced gluon radiation spectrum to, respectively, first and second order in opacity. We discuss which new medium-induced color correlations arise, how the weight of these contributions depends on the relation between the momentum and length scales in the problem (i.e. on the formation times), and how higher orders in opacity enhance the probability that the radiated gluons are decohered in color from the leading partonic fragments. 
The readers who wonder more about the conceivable effects of medium-modified color flow on jet observables, but are less interested in the technical aspects of our calculation, may jump in a first reading directly to Section~\ref{sec5}.
There, we investigate how the output of standard hadronization models will change if they are interfaced with the characteristic medium-modified color flow patterns identified in our calculations of sections~\ref{sec3} and ~\ref{sec4}. Much of this discussion will focus on the \textsc{\textsc{Lund}} string fragmentation model, but we discuss also implications for cluster hadronization models.
Sec.~\ref{sec5}, read together with Sec.~\ref{sec2}, provide a sufficiently self-contained digest of our qualitative arguments and of the essential message conveyed by our paper.  
We finally summarize and discuss our findings and provide an outlook to further open questions. 
\section{Medium modification of parton splittings and their color connections}
\label{sec2}
\begin{figure}[!htp]
\begin{center}
\includegraphics[clip,width=0.4\textwidth]{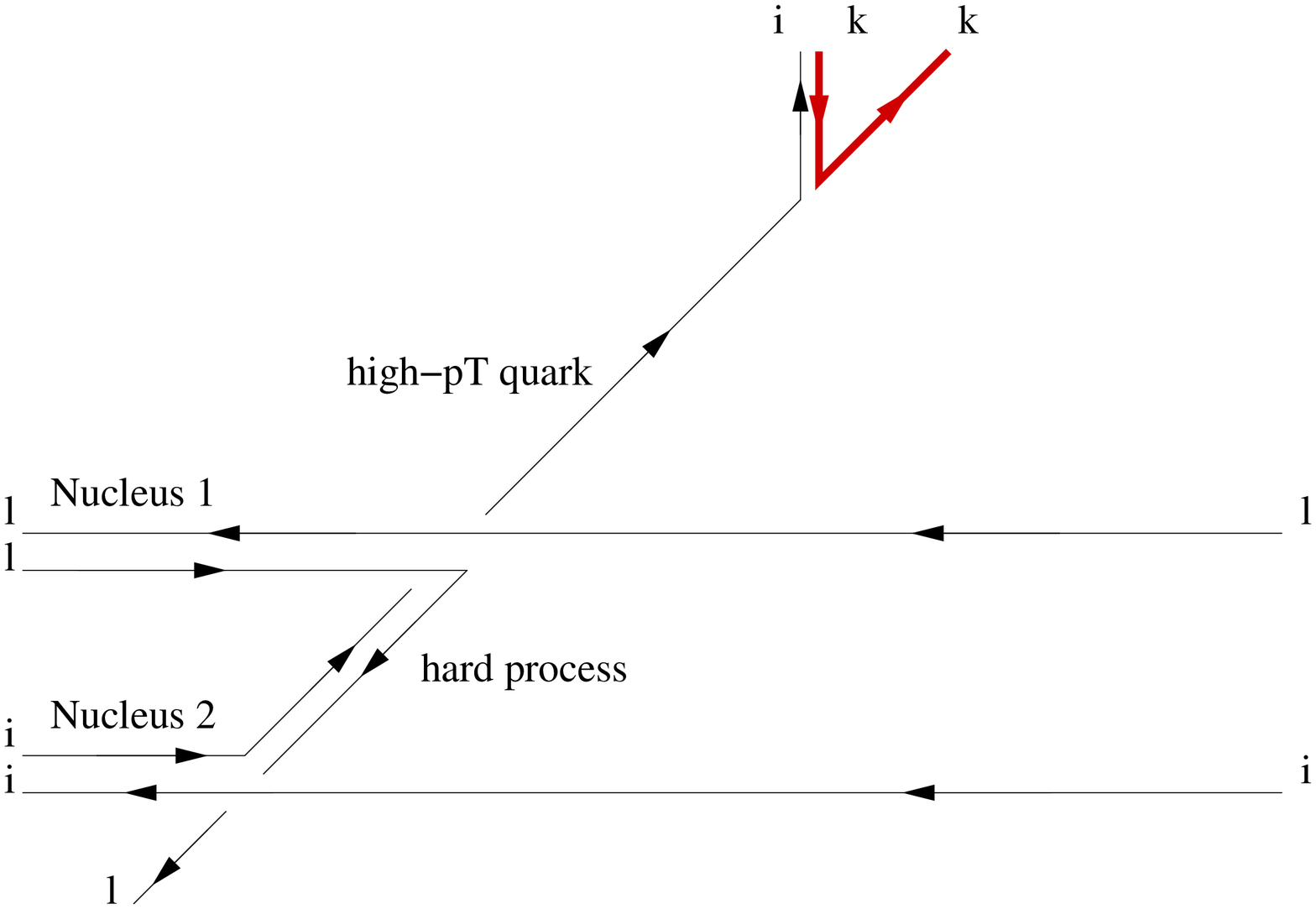}
\hskip 1cm
\includegraphics[clip,width=0.4\textwidth]{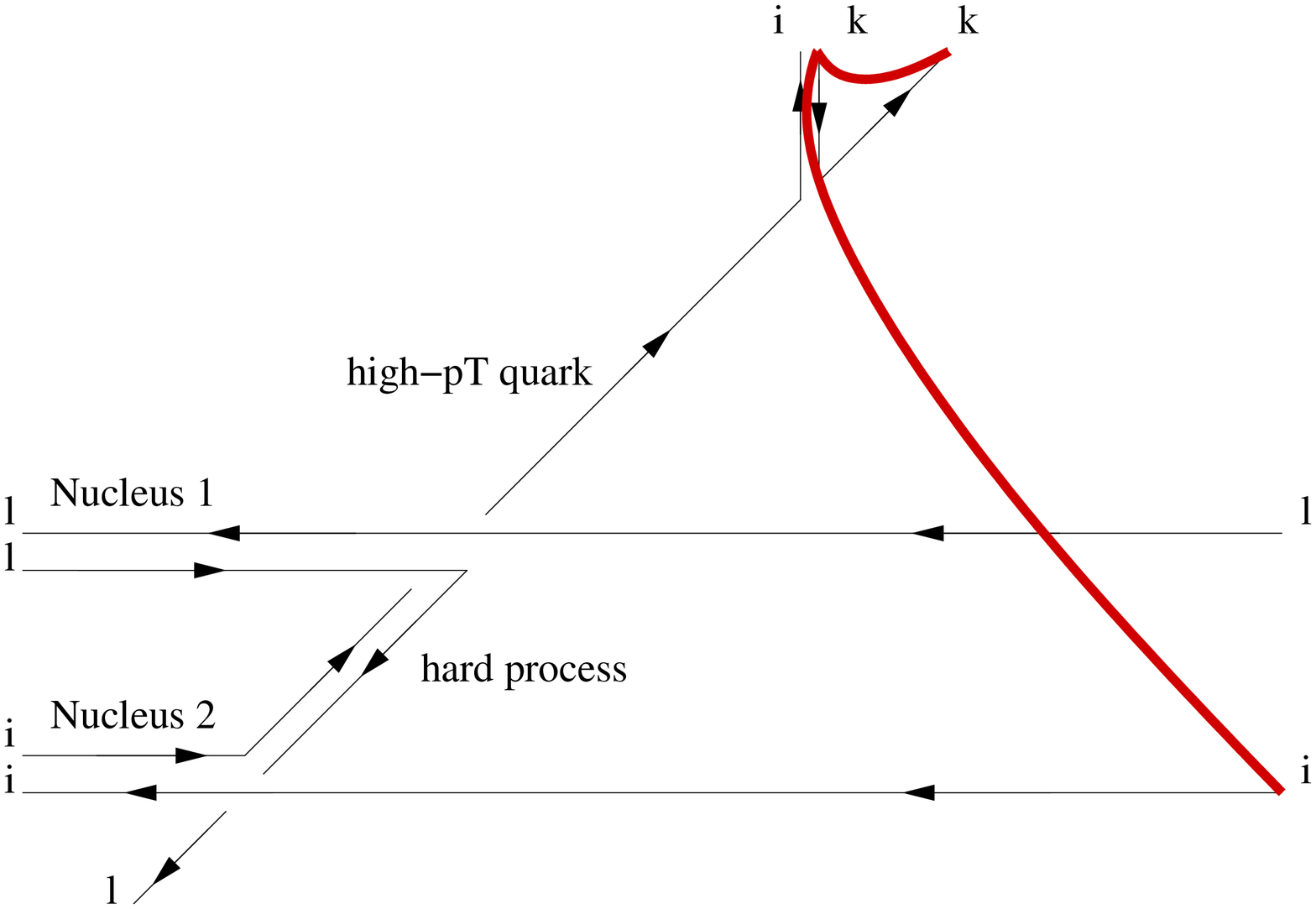}
\caption{The vacuum baseline: a hadron-hadron collision with a hard process followed by a $q\to qg$ splitting in the shower stage. Color-flow in the large-$N_c$ limit is shown explicitly and the color-singlets (a cluster in the left panel, a string in the right panel) to be interfaced with the hadronization routine (\textsc{Herwig} cluster-decay or \textsc{Pythia} string-fragmentation, respectively) are displayed in color.} 
\label{fig1}
\end{center}
\end{figure}

To set the stage for the study of medium-modifications of high-$p_T$ processes to be carried out in this paper, we start from the `vacuum' baseline of a hard partonic interaction in the absence of medium effects.
To be specific, we first consider the case of a quark of color $l$ from `hadron 1' (proton/nucleus) that hard scatters on a quark of color $i$ from `hadron 2' (proton/nucleus) and subsequently radiates a gluon in the showering stage. To leading order in the coupling constant $\alpha_s$ (when the hard scattering occurs via a single gluon exchange) and to leading order in the number of colors $N_c$ (when gluons can be represented as quark-anti-quark pairs), this vacuum process is depicted in Fig.~\ref{fig1}.
In order to ensure color neutrality, the colliding hadrons -- here and in the following -- will be depicted schematically as two opposite color charges. 
In the following sections, we shall calculate medium-modifications of such color-differential partonic processes and discuss how the medium-modified color flow affects their hadronization.
Color connections are shown schematically in Fig.~\ref{fig1} by  supplementing the quarks of color $l$ and $i$ in the incoming nuclei with anti-quarks of the corresponding color. These anti-quarks  should be thought of as formal placeholders for what remains of a hadron once a quark of color $l$ or $i$ has been taken from it. They do not partake in the partonic interaction, but provide the color reservoir with which the scattered quarks are color correlated.  

The dynamics underlying hadronization is not understood from
first principles. It is known, however, that the color flow of the underlying perturbative process is relevant and this is implemented in modern, phenomenologically successful models of hadronization, such as the \textsc{Lund} string-fragmentation in \textsc{Pythia} or the cluster-decay in \textsc{Herwig}. Here we discuss shortly how information about the color flow in the partonic process enters these models.

Cluster hadronization models, as implemented e.g. in the \textsc{Herwig} event generator \cite{Webber:1983if}, group  the result of a perturbative shower evolution into a set of color singlet clusters  by splitting each gluon in the final state into a $q\bar{q}$-pair;
clusters are then decayed and hadronized independently. This is illustrated in Fig.~\ref{fig1} (left), where the quark $k$ is combined with the anti-quark of the split gluon into a cluster. The most energetic hadron is then typically a fragment of this cluster, and the distribution of the fragments will depend on the momentum of the cluster and on its invariant mass. 
Alternatively, the \textsc{Lund} model, as implemented e.g. in the  \textsc{Pythia} event generator \cite{Sjostrand:2006za}, groups the same perturbative information into a set of color strings that start with a quark, follow the color flow by including gluons as kinks, and end on an anti-quark; these color singlet strings are then hadronized (through excitation of $q\bar{q}$ pairs from the vacuum) according to a prescription that
requires kinematic information about both the end-points of the string and all the kinks. The multiplicity and distribution of final hadrons will then depend significantly on the `length of the string', that is on the separation of the quark and anti-quark end-points in momentum space. 

The presence of a medium with which high-energy partons can exchange color can clearly alter the color connections  described above and, by changing the properties of the clusters/strings, have an effect on the final hadron spectra.
In the remainder of this section we illustrate the essential ideas, focusing on the simple situation of a single interaction of the hard parton with the medium.   We treat the cases of an incoming quark and gluon separately.
\begin{figure}[!htp]
\begin{center}
\includegraphics[clip,width=0.4\textwidth]{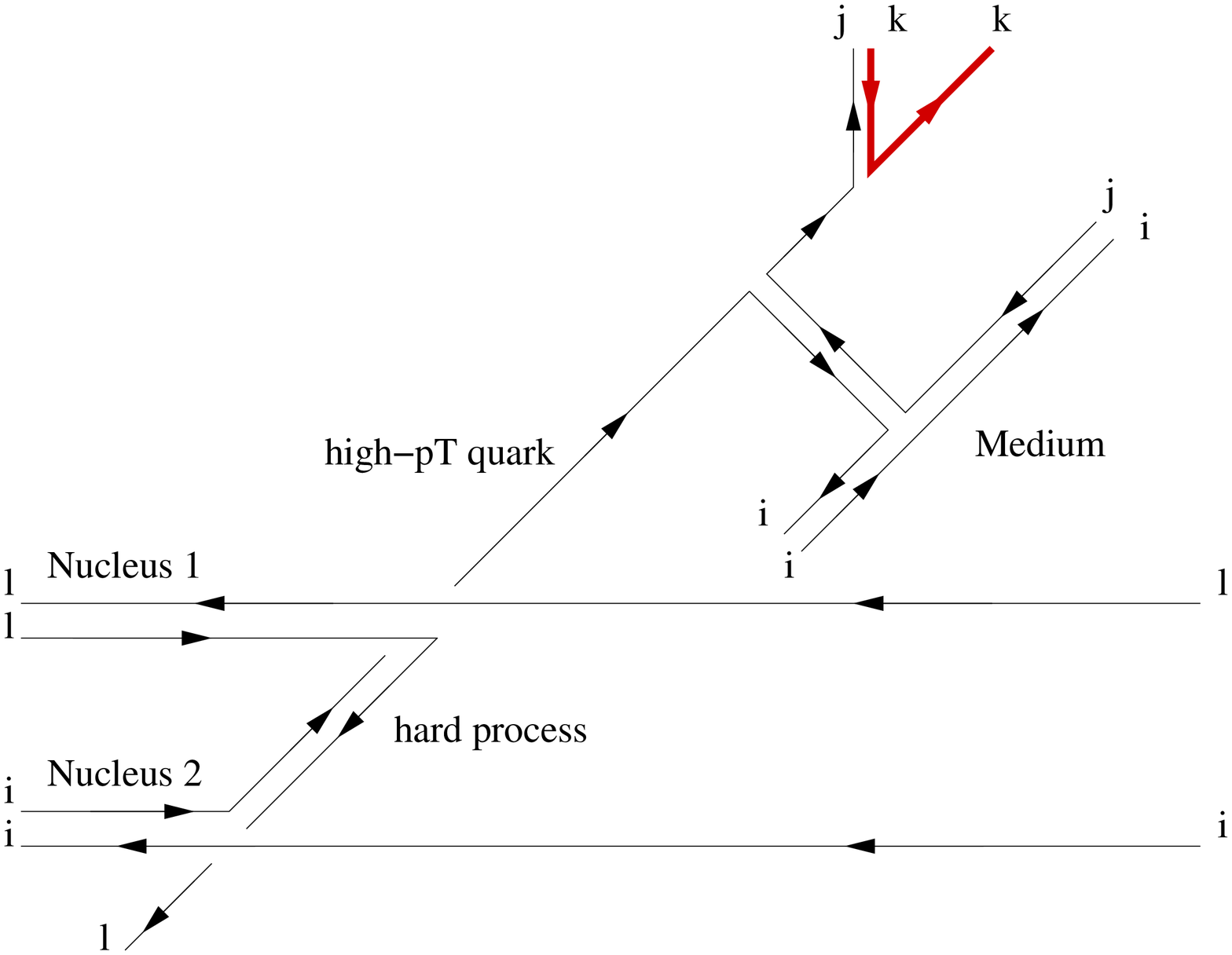}
\hskip 1cm
\includegraphics[clip,width=0.4\textwidth]{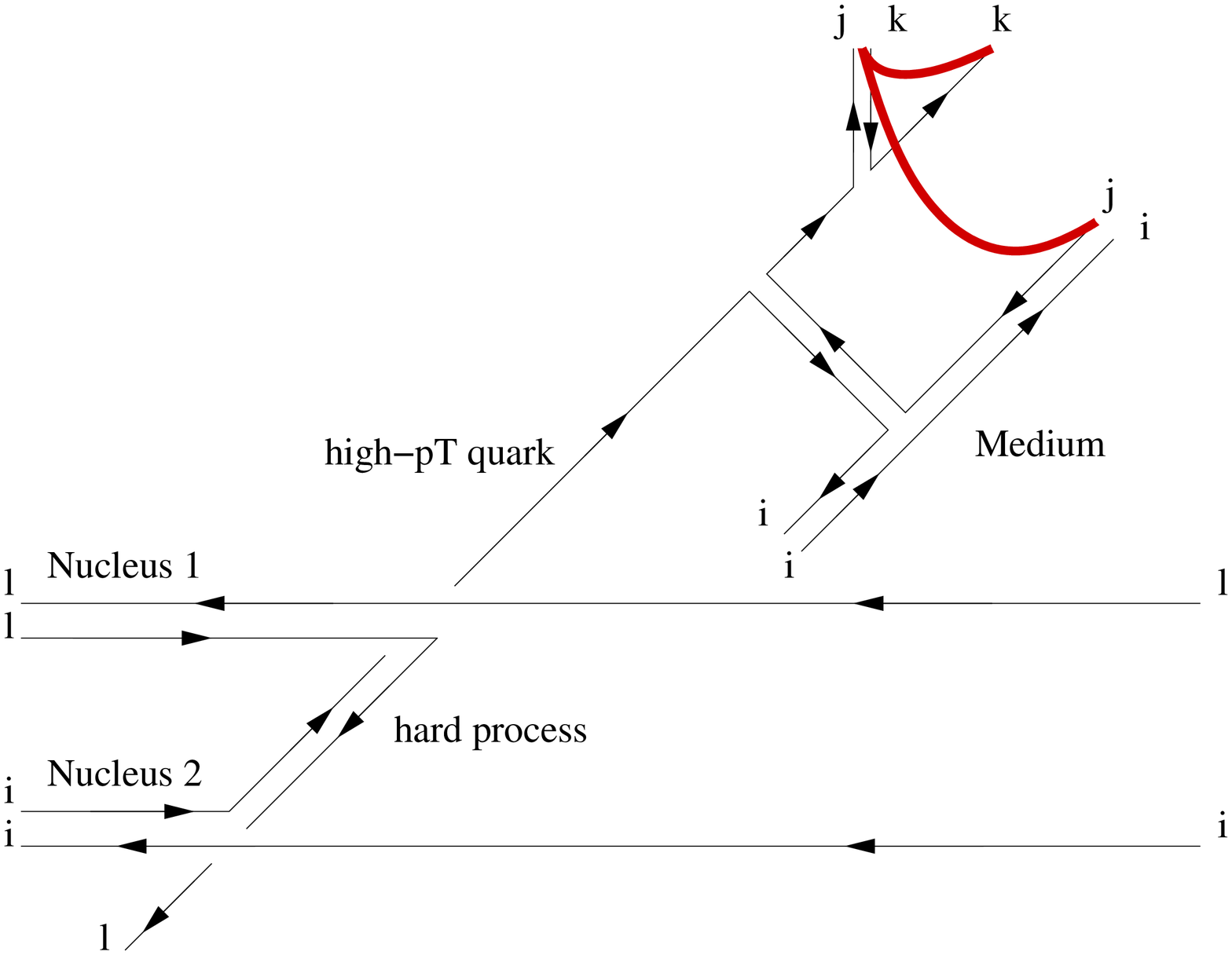}
\caption{Example of a hard $q\, q \to q\, q$ event embedded in a nucleus-nucleus collision in which one of the high-$p_T$ quarks interacts once with the surrounding QCD matter which induces gluon radiation. Gluons are denoted by $q\bar{q}$-pairs. The red lines denote the color singlet into which the leading quark $k$ is grouped to form a cluster (left-hand side) or a \textsc{Lund}-string (right-hand side) in the corresponding hadronization models.} 
\label{fig2}
\end{center}
\end{figure}

\subsection{Medium-induced color flow for a quark projectile to first order in opacity}
\label{sec2a}
\begin{figure}[!htp]
\begin{center}
\includegraphics[clip,width=0.4\textwidth]{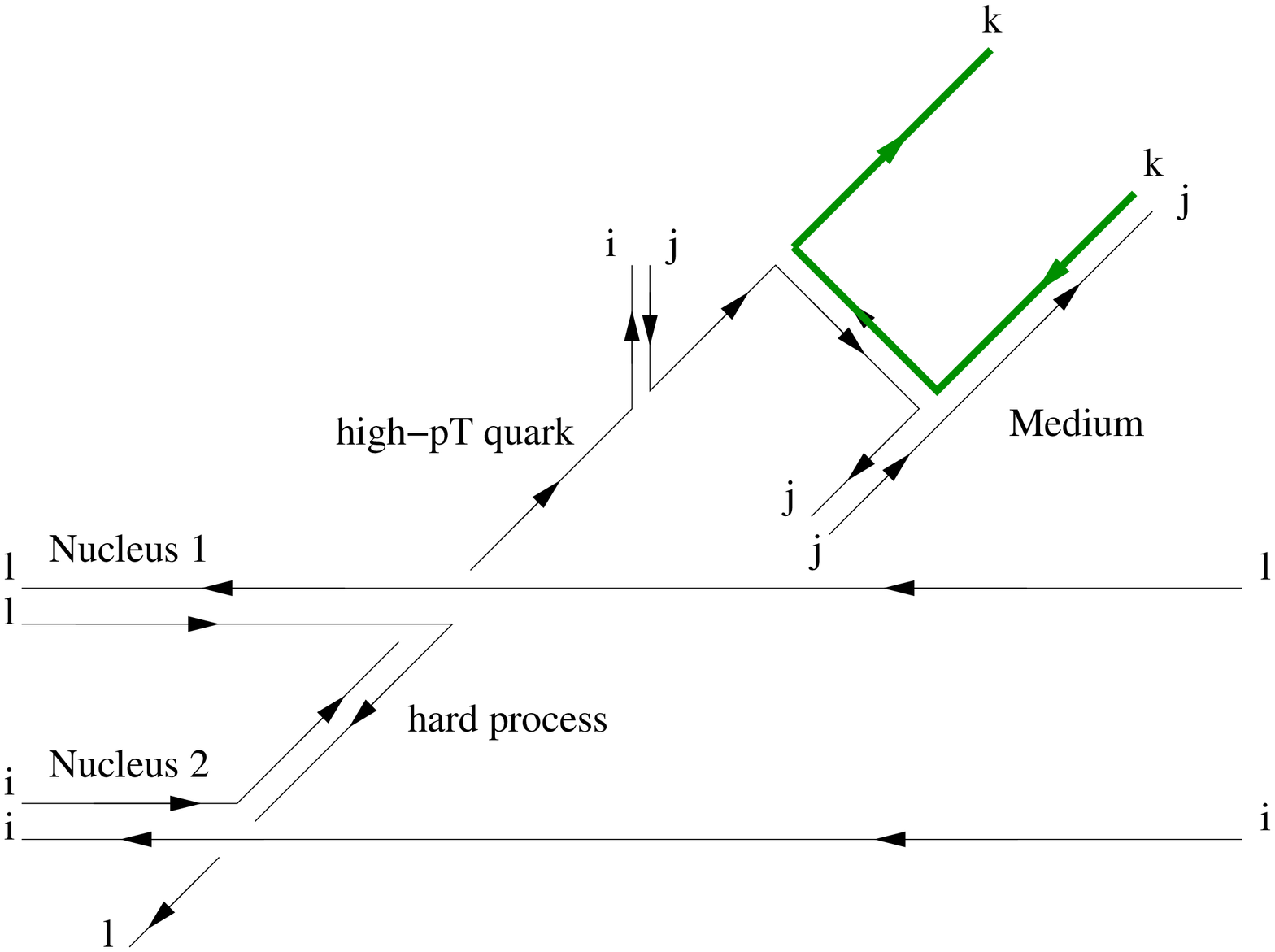}
\hskip 1cm
\includegraphics[clip,width=0.4\textwidth]{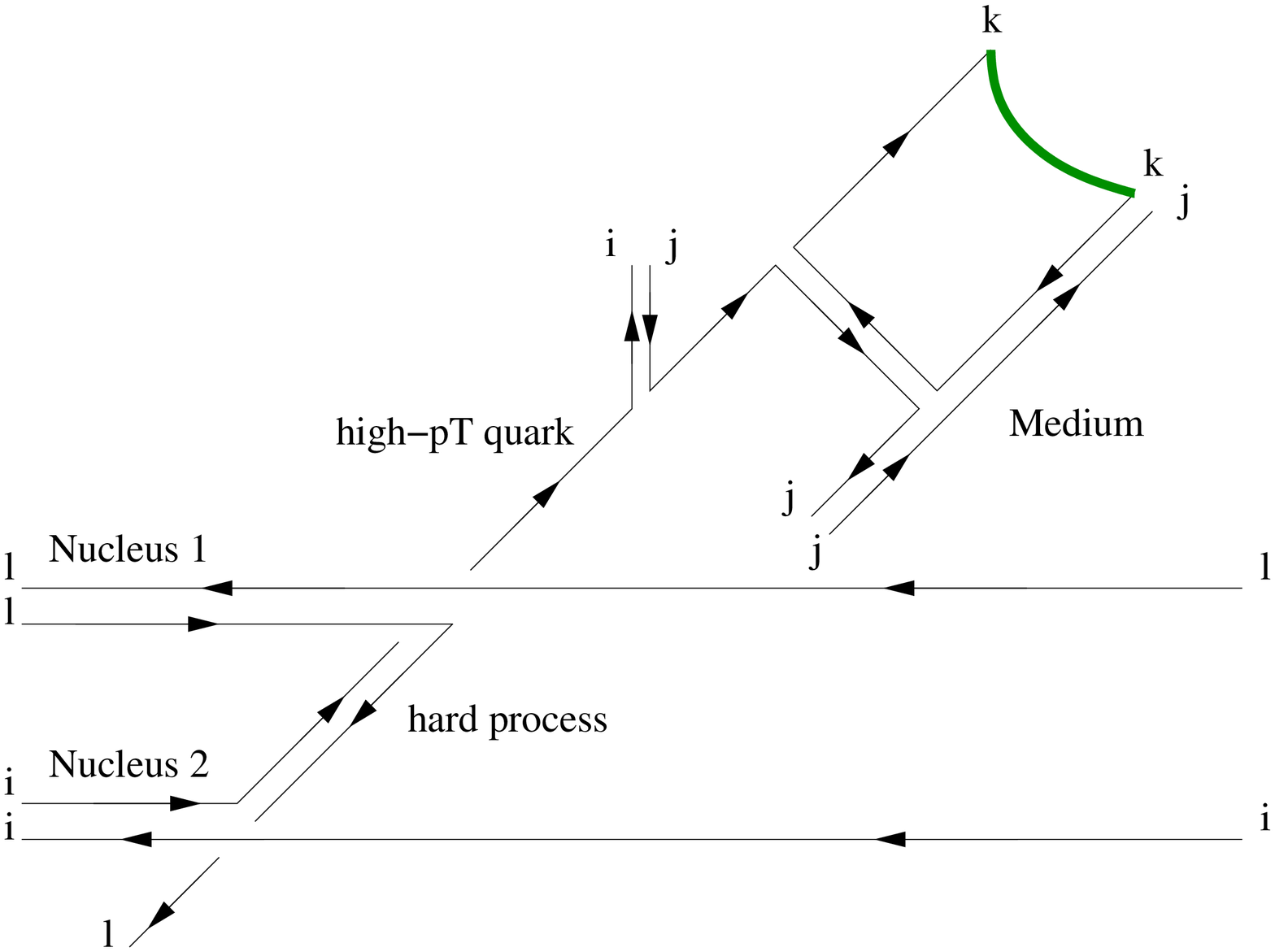}
\caption{Same as Fig.~\ref{fig2}, but for the case that the high-$p_T$ quark radiates the gluon prior to interacting with the medium.} 
\label{fig3}
\end{center}
\end{figure}

`Jet-quenching' calculations consider the interaction of high-energy partons in the dense QCD medium produced in heavy ion collisions. 
In Figs.~\ref{fig2} and~\ref{fig3}, we display the simplest case of such an interaction: an elastic scattering of a high-energy quark in the plasma induces the radiation of a gluon. The interaction between the quark projectile and the medium occurs through the exchange of a single gluon,  depicted here as a $q\bar{q}$-pair in the large $N_c$-limit. 

For the configuration in Fig.~\ref{fig2}, the `leading' color singlet cluster, the one containing the quark $k$, shows the same color structure 
as the vacuum baseline, the color-flow connecting the high-$p_T$ quark with the anti-quark component of the gluon. 
In the \textsc{Lund} picture, the string containing the leading quark $k$ links to the radiated gluon as in the vacuum baseline in Fig.~\ref{fig1}, but now ends 
on an anti-quark from the medium. However,  similarly to the vacuum case, this end point sits at low transverse momentum (with respect to the energy of the hard parton). For the above reasons, we shall refer to the color configuration depicted in Figure~\ref{fig2} as {\it vacuum-like} or -- emphasizing the link of the radiated gluon with the projectile fragment -- {\it projectile-connected}.  

Figure~\ref{fig2} is only one possible color structure that can emerge from a single interaction of the projectile with the QCD medium. The second possibility is shown in Fig.~\ref{fig3}, where (from the point of view of color flow) interaction with the medium occurs after the gluon emission. 
As a consequence, the leading color singlet cluster combines a quark at projectile energy with a target component at low (thermal) $p_T$. In \cite{Beraudo:2011bh} the invariant mass of this cluster was shown to be parametrically larger than the one of the cluster in Fig.~\ref{fig2}. Analogously, in the \textsc{Lund} framework the leading string connects the quark $k$ directly to the target. The radiated gluon is, in both descriptions, color decohered from the projectile and will contribute only to an increase of the multiplicity of soft hadrons. In the following, we shall refer to these color configurations as {\it medium-modified} or {\it gluo-decohered}. 

\begin{figure}[!tp]
\begin{center}
\includegraphics[clip,width=1.0\textwidth]{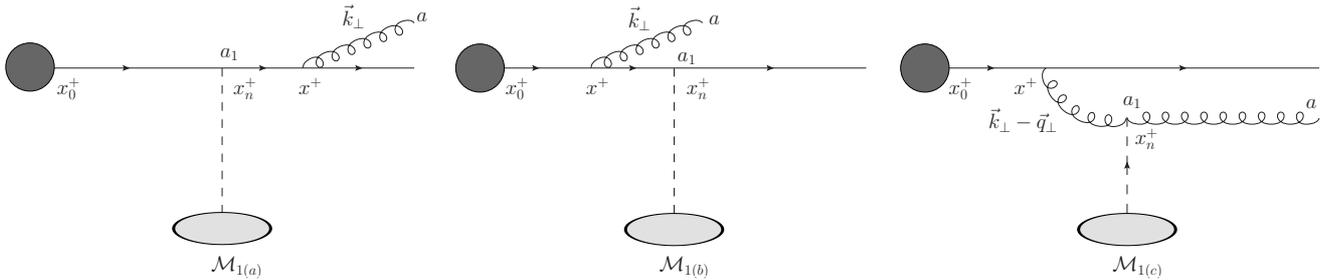}
\caption{The ($N\!=\!1$ in opacity) gluon-radiation amplitude arising from a single interaction of the projectile with the scattering centers of the medium. 
} 
\label{fig4}
\end{center}
\end{figure}

We finally relate this discussion to the diagrams in Fig.~\ref{fig4} that are usually drawn for the calculation of parton energy loss within an opacity expansion. In the figure, the initial hard process is sub-summed in a dark blob from which a single high-$p_T$ quark line emerges at the light-cone time $x_0^+$ (employing light-cone coordinates will be convenient in the following). 
The interaction of the high-$p_T$ quark with the medium is described in terms of one-gluon exchange with a colored scattering center, or in terms of multiple such one-gluon exchanges.
There is a direct correspondence between the Feynman diagrams in Fig.~\ref{fig4} and the ones displaying the color-flow in the large-$N_c$ limit in Figs.~\ref{fig2} and~\ref{fig3}.
One easily checks that the first amplitude in Fig.~\ref{fig4} corresponds 
to the color flow shown in Fig.~\ref{fig2}, while the second diagram refers to the color configuration depicted in Fig.~\ref{fig3}.
Finally, the amplitude with the triple gluon vertex contributes to both color channels shown in Figs.~\ref{fig2} and ~\ref{fig3} (for details, see section~\ref{sec3}).
This correspondence prompts us to label the two color channels as Final State Radiation (FSR) and Initial State Radiation (ISR), depending on whether the gluon radiation occurs after or before the elastic scattering.

A detailed color-differential calculation of medium-induced gluon radiation starting from the diagrams in Fig.~\ref{fig4} will be presented in Sec.~\ref{sec3}.
\subsection{Medium-induced color flow for a gluon projectile to first order in opacity}
\label{sec2b}
\begin{figure}[!h]
\begin{center}
\includegraphics[clip,width=0.35\textwidth]{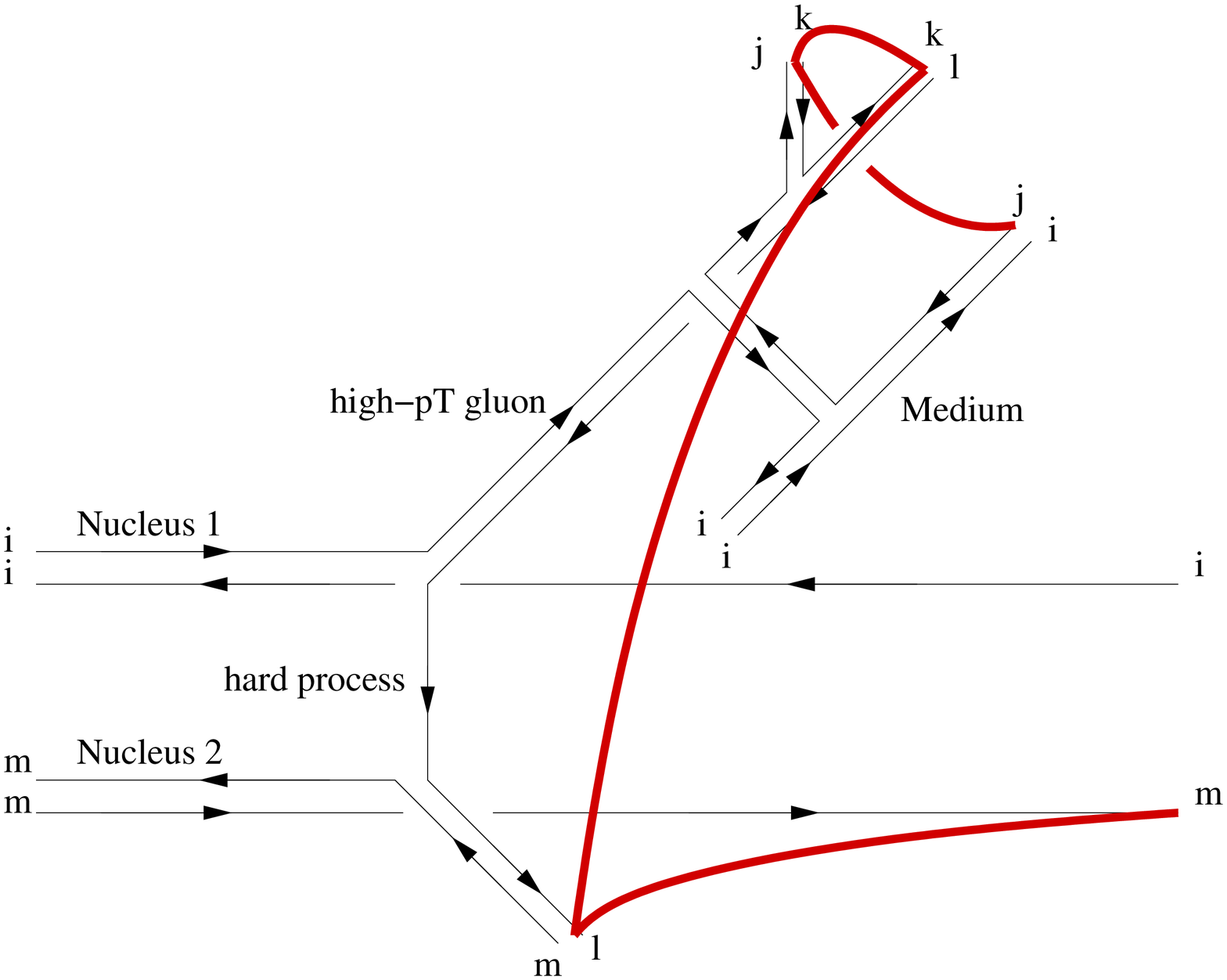}
\hskip 1cm
\includegraphics[clip,width=0.35\textwidth]{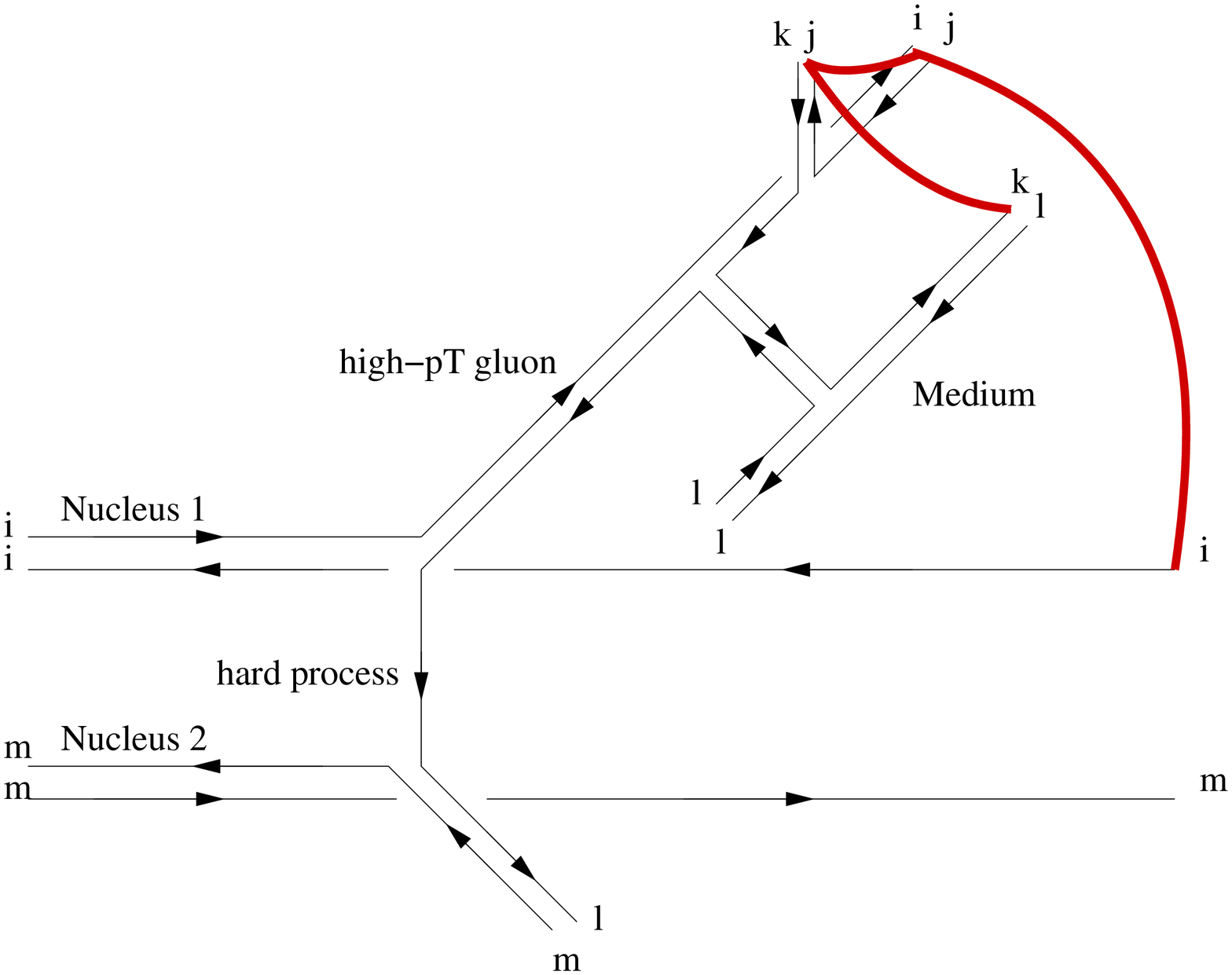}
%
%
\includegraphics[clip,width=0.35\textwidth]{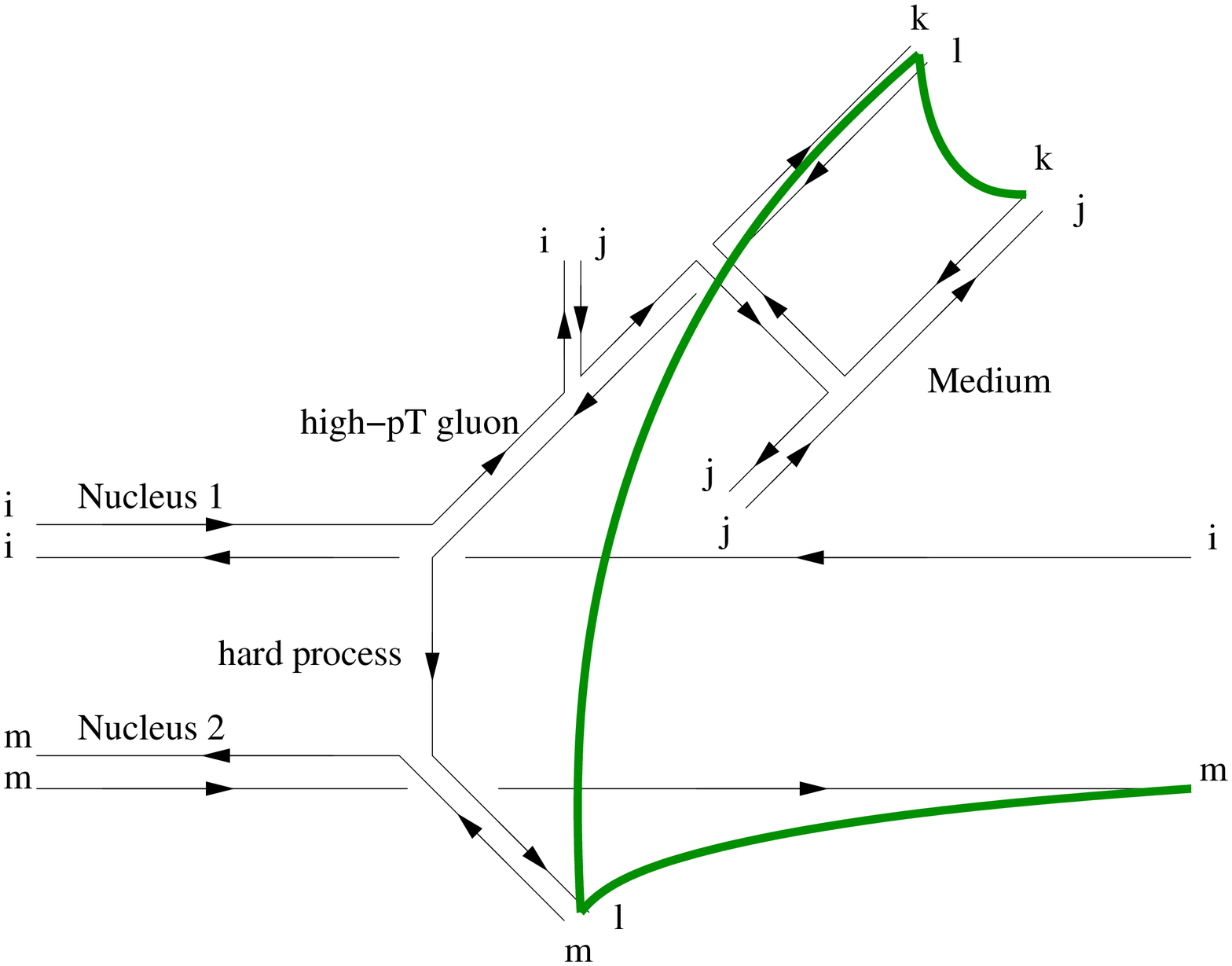}
\hskip 1cm
\includegraphics[clip,width=0.35\textwidth]{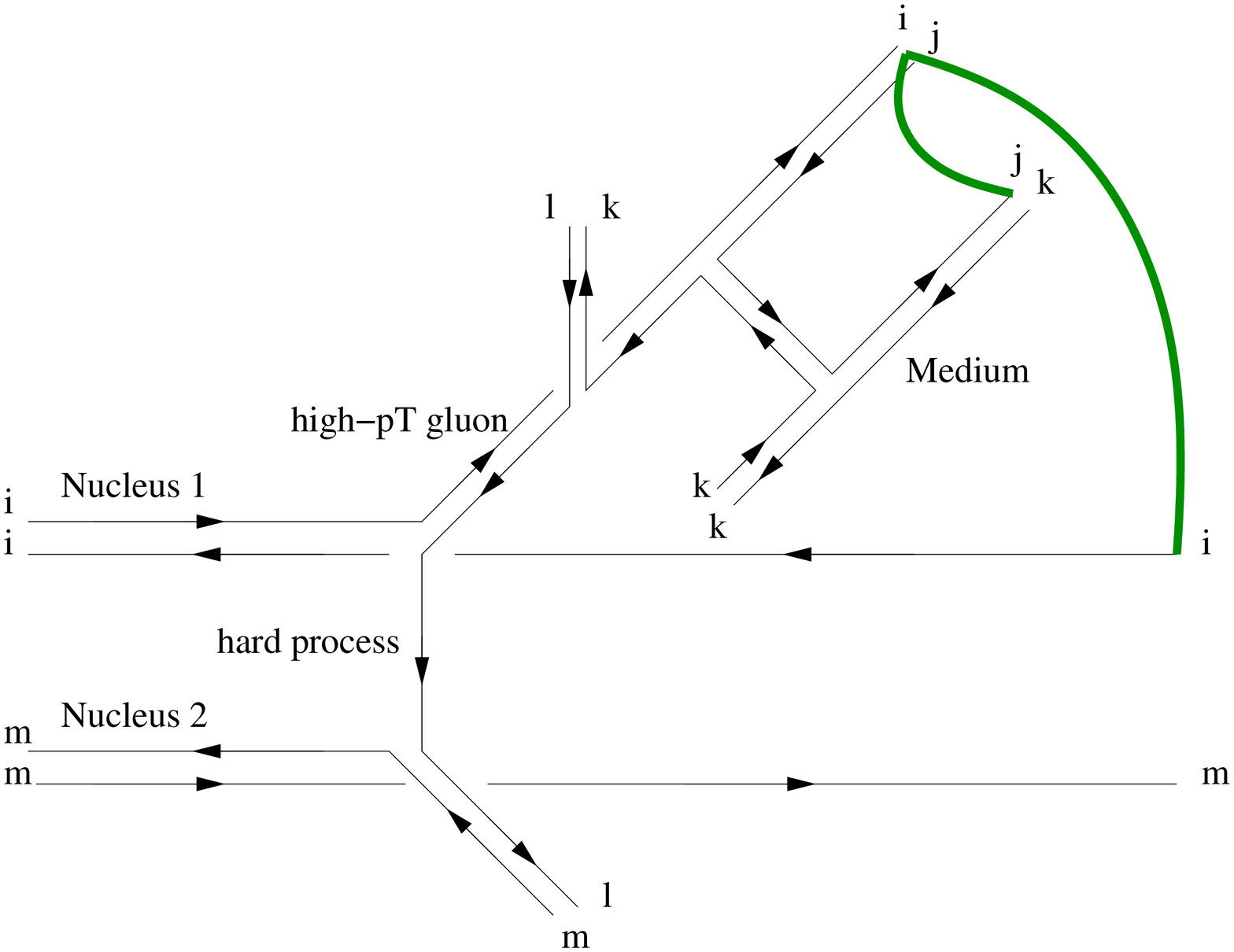}
%
\includegraphics[clip,width=0.35\textwidth]{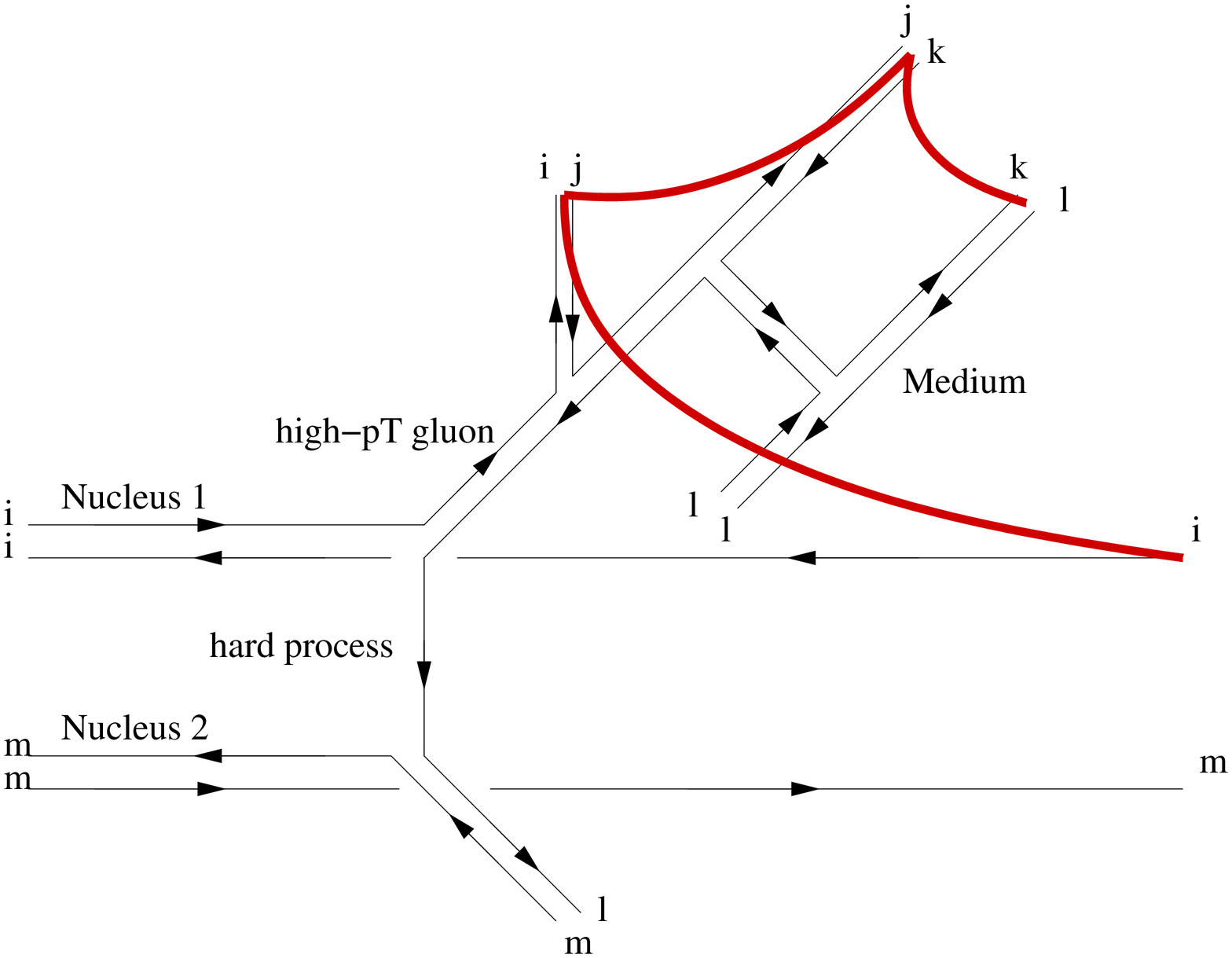}
\hskip 1cm
\includegraphics[clip,width=0.35\textwidth]{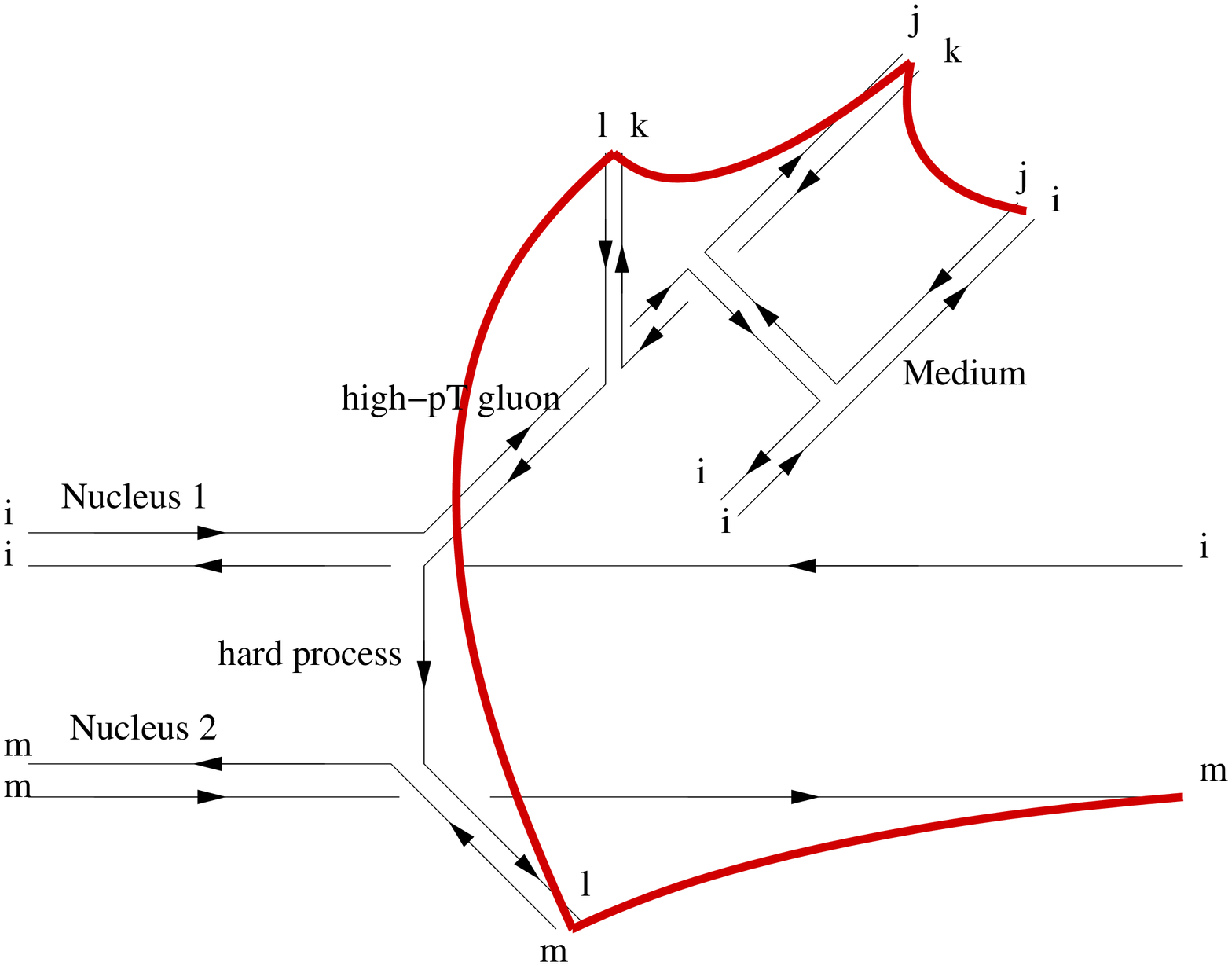}
\caption{The six different color configurations that arise to first order in opacity for a high-$p_T$ gluon. 
Curved colored lines represent the `leading' \textsc{Lund} strings, i.e. the ones connected to the most energetic parton. 
Representing the high-$p_T$ gluon as a $q\bar{q}$-pair, there are 4 contributions corresponding to scattering + Final State Radiation on the same quark or anti-quark leg (diagrams in first row) and to the corresponding terms for the Initial State Radiation case (diagrams in second row): in these four channels the second component of the gluon acts as a spectator.
The contributions in the third row for which the radiated gluon is emitted from the $q$-leg while the medium couples to the $\bar{q}$-leg of the high-$p_T$ gluon (or vice-versa)
are suppressed in the soft ($x_g\ll1$) limit.} 
\label{fig5}
\end{center}
\end{figure}

For the case of a high-$p_T$ gluon produced in a hard scattering process, a larger number of color channels is involved. In Figure~\ref{fig5} we show the color configurations that arise to first order in opacity~\footnote{Here and in the following, we distinguish the two daughter gluons in the medium-modified $g \to g\, g$ splitting as leading (i.e. carrying $1-x_g\sim1$) and 
subleading (i.e. carrying $x_g \ll 1$), respectively. Exchanging these labels of the two daughter gluons would amount to exchanging the diagrams on the left and right column of 
Fig.~\ref{fig5}.}. For each case, we sketch only the associated \textsc{Lund} string that contains the most energetic gluon.

In the large-$N_c$ limit, a gluon is represented as a quark-anti-quark pair. There are contributions to medium-induced parton branching in which both the scattering and the radiation occur off the $q$  ($\bar{q}$) leg of the projectile gluon while the $\bar{q}$ ($q$)  leg is a silent spectator. In close analogy to the case of a quark projectile, we distinguish four cases in which -- \emph{from the point of view of color flow} -- the gluon branching occurs either on the quark or on the anti-quark leg, and either after (Final State Radiation) or before (Initial State Radiation) color exchange with the medium. We refer to these four cases as FSR$(q)$, FSR$(\bar{q})$, ISR$(q)$, ISR$(\bar{q})$. In addition, there are two channels in which the gluon radiation is connected, in the large-$N_c$ limit, to the $q$-leg ($\bar{q}$-leg) of the high-$p_T$ gluon while color-exchange with the medium occurs via the $\bar{q}$-leg ($q$-leg), see last row of Fig.~\ref{fig5}.
We anticipate that these last two channels will provide a subleading contribution to the spectrum in the case of soft ($x_g\ll 1$) gluon radiation.
An explicit evaluation of the color-differential radiation spectrum will be presented and discussed further in Sec.~\ref{sec3} and in appendix~\ref{appa}.
Here, we note that the \textsc{Lund} string including a high-$p_T$ gluon will inevitably have both end-points at (anti)quarks that typically carry low transverse momentum. It is a particularly `long' string in the sense that it stretches from low-$p_T$ to high-$p_T$ and back again.

\section{Color flow and radiation spectrum in the $N=1$ opacity expansion}
\label{sec3}
In this section, we provide a QCD-based color differential calculation of the medium induced gluon radiation process discussed in Sec.~\ref{sec2}.
We do so within the framework of parton energy loss first formulated by Baier, Dokshitzer, Mueller, Peign\'e and Schiff (BDMPS) \cite{Baier:1996sk}. The QCD medium is parametrized, following the Gyulassy-Wang model \cite{Gyulassy:1993hr}, as a collection of static scattering centers at a discrete set of space time points $x_n$ giving rise to a potential $A^\mu(x)$. We work in light-cone coordinates and in a high-energy approximation in which (in the light-cone gauge $A^+=0$) only the `$-$' light-cone component for the scattering potential is relevant
\beq
A^-(x)\equiv\sum_{n=1}^{N}\int\frac{d\q}{(2\pi)^2}e^{i\q\cdot(\x-\x_n)}{\mathcal A}(\q)\;\delta(x^+\!-\!x_n^+)\;T_{(n)}^{a_n}\otimes T_{(R)}^{a_n} \, .
\eeq
Here, $R$ denotes the representation of the parton suffering the elastic scattering. 
It is customary to formulate calculations of parton energy loss in a rotated frame in which the longitudinal axis points along the initial direction of propagation of the high transverse momentum parton. This parton has thus a large initial longitudinal momentum and no initial transverse momentum.
During its in-medium propagation it will accumulate transverse momentum $\q$ from interactions with the medium, and will lose a light-cone energy fraction $x_g$ by emitting a gluon of transverse momentum $\kkg$.
For a parton of mass $M$ and virtuality $Q$, the incoming and outgoing momenta read 
\beq
p_i=\left[p^+,\frac{M^2+Q^2}{2p^+},\0\right],\qquad p_f=\left[(1-\xg)p^+,\frac{(\q-\kkg)^2+M^2}{2(1-\xg)p^+},\q-\-\kkg\right]\, ,
\eeq
and for the gluon four-momentum and polarization vector one has
\beq
\kg=\left[\xg p^+,\frac{\k_g^2}{2\xg p^+},\k_g\right],\qquad \eg=\left[0,\frac{\eeg\cdot\kkg}{\xg p^+},\eeg\right]\, .
\eeq
To illustrate the essential ideas within a sufficiently simple setup, we calculate in this section explicit color differential expressions for the gluon radiation at $N=1$ order in the opacity expansion. 
\subsection{Color-differential gluon radiation off a quark: the case $N=1$, $x_0^+=-\infty$}
\label{sec3A}
\begin{figure}[!tp]
\begin{center}
\includegraphics[clip,width=1.0\textwidth]{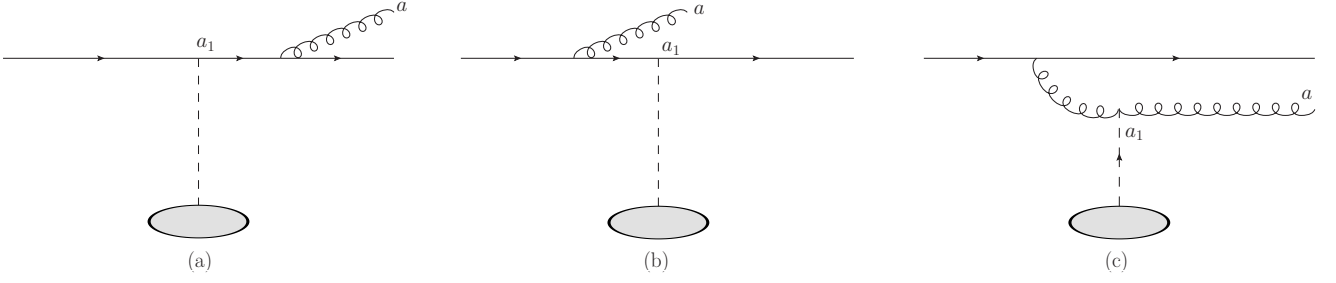}
\caption{The three diagrams contributing to the gluon radiation off an on-shell parton suffering a single scattering in an external color field.} 
\label{fig6}
\end{center}
\end{figure}

We consider first the particularly simple case of a high-energy quark produced in the distant past ($x_0^+ = - \infty$) and that interacts just once with the medium.  The diagrams contributing to this process are shown in Fig.~\ref{fig6}.
The amplitude corresponding to gluon emission after the scattering (first diagram in Fig.~\ref{fig6}) takes the simple form
\beq\label{eq:inf_after}
i{\cal M}_{(a)}
=-ig\, (t^at^{a_1})\sum_{n=1}^{N}\frac{p_f\!\cdot\!\eg}{p_f\!\cdot\!\kg}\,(2p^+) {\cal A}(\q)\,e^{iq\cdot x_n}\,T_{(n)}^{a_1}\, .
\eeq
Analogously, one can write the expressions for the other two Feynman diagrams shown in Fig.~\ref{fig6}. The sum of these three contributions can be written in a compact form
\beqa
i{\cal M} &=& i{\cal M}_{(a)}  + i{\cal M}_{(b)}  + i{\cal M} _{(c)} \nonumber\\
  &=&  -2ig\, [t^a,t^{a_1}]\sum_{n=1}^{N} \left[\frac{\eeg\!\cdot\!\kkg}{\kkg^2\!+\!\xg^2M^2}-\frac{\eeg\!\cdot\!(\kkg\!-\!\q)}{(\kkg\!-\!\q)^2\!+\!\xg^2M^2}\right]
 \,(2p^+) {\cal A}(\q)\,e^{iq\cdot x_n}\,T_{(n)}^{a_1}\,.
 \label{mGB}
\eeqa
%
%
Here and in the following, we work in the soft ($x_g \ll 1$) limit in which terms subleading in $x_g$ are neglected. However, we have kept in Eq.~(\ref{mGB}) terms of order $\xg^2\, M^2$, so that the discussion can readily be extended to deal with the heavy-flavor case.
The color-inclusive spectrum of radiated gluons is obtained from $\vert {\cal M}\vert^2$ after averaging (summing) over the initial (final) states:
\beq
\kg^+\frac{dN_g}{d\kkg d\kg^+}\equiv
\frac{1}{\sigma^{\rm el}}\kg^+\frac{d\sigma^{\rm rad}}{d\kkg d\kg^+}
= C_A\frac{\alpha_s}{\pi^2}\left\langle\left[\K_0-\K_1\right]^2\right\rangle \, ,\label{eq:gb}
\eeq
where the notational shorthands 
\beq
\K_0\equiv\frac{\kkg}{\kkg^2\!+\!\xg^2M^2},\qquad \K_1\equiv\frac{\kkg\!-\!\q}{(\kkg\!-\!\q)^2\!+\!\xg^2M^2}\, ,
\eeq
have been used, and the momentum kicks $\q$ received from the medium were averaged according to the corresponding elastic cross section:
\beq
\big\langle\dots\big\rangle\equiv\int d\q\,\left(\frac{1}{\sigma^{\rm el}}\frac{d\sigma^{\rm el}}{d\q}\right)(\dots)
=\int d\q\, \vert {\cal A}(\q)\vert^2 (\dots)\, .\label{eq:aver}
\eeq
In the $M\to 0$ limit, the well-known Gunion-Bertsch spectrum
\beq
\kg^+\frac{dN_g}{d\kkg d\kg^+}=C_A\frac{\alpha_s}{\pi^2}
\left\langle\frac{\q^2}{\kkg^2(\kkg\!-\!\q)^2}\right\rangle\, ,
\label{gunionbertsch}
\eeq
is recovered.
Notice that once the radiation spectrum is normalized by its elastic cross section, the overall color factor $C_A$ is universal, irrespective of  whether the incoming projectile is a quark or a gluon.
The elastic scattering cross section entering Eq.~(\ref{eq:aver}) is often chosen to be of Yukawa-type, $\vert {\cal A}(\q)\vert^2 \propto {\mu_D^2}/{\left( \q^2 + \mu_D^2\right)^2}$, but we shall not rely on a specific functional shape
of $\vert {\cal A}(\q)\vert^2$ in the following.

Extending the derivation of the Gunion-Bertsch spectrum to the color-differential case is straightforward. After identifying in the three-gluon vertex $\propto[t^a,t^{a_1}]$ 
the two different `color-orderings', one writes the total radiation amplitude as:
\beq
i{\cal M} = i{\cal M}^{a\, a_1}+i{\cal M}^{a_1\, a}\, , 
 \label{mGBdiff}
\eeq
where ${\cal M}^{a\, a_1}\!\propto\! t^a t^{a_1}$ and ${\cal M}^{a_1\, a}\!\propto\! t^{a_1} t^a$, respectively. One then notes that in the cross section the interference between the two contributions is suppressed by a factor $1/N_c^2$, since 
${\rm Tr}(t^at^{a_1}t^{a_1}t^a)=C_F^2N_c$ and ${\rm Tr}(t^at^{a_1}t^at^{a_1})=-({1}/{2N_c})C_FN_c$. Thus,
\beq
\sigma^{\rm rad}=\sigma_{aa_1}^{\rm rad}+\sigma_{a_1a}^{\rm rad}+\mathcal{O}(1/N_c^2)\, ,
\eeq
from which one obtains, to leading order in $N_c$, the spectrum of radiated gluons for each color channel. Keeping track of the exact dependence on the energy fraction carried by the emitted gluon one finds
\beqa
\label{eq:aa1inf}
\left.\kg^+\frac{dN_g}{d\kkg d\kg^+}\right|_{aa_1}&=&\frac{N_c}{2}\frac{\alpha_s}{\pi^2}\left\langle\left[\overline{\K}_0-\K_1\right]^2\right\rangle\, , \\
\label{eq:a1ainf}
\left.\kg^+\frac{dN_g}{d\kkg d\kg^+}\right|_{a_1a}&=&\frac{N_c}{2}\frac{\alpha_s}{\pi^2}\left\langle\left[\K_0-\K_1\right]^2\right\rangle\, ,
\eeqa
where we have used the shorthand $\overline{\K}_0\equiv\frac{\kkg\!-\!\xg\q}{(\kkg\!-\!\xg\q)^2\!+\!\xg^2M^2}$.  To leading order in $\xg$ (i.e. in the soft limit), one sees that both color channels have exactly the same weight and that the sum of  the two contributions yields Eq.~(\ref{eq:gb}). 
In general, however, there is no general argument for why different color channels should contribute equally. To illustrate this point already here, we have kept in $\overline{\K}_0$ the subleading dependence on $\xg$.
\subsection{Color-differential gluon radiation off a gluon: the case $N=1$, $x_0^+=-\infty$}
\label{sec3b}
So far, the calculations in this section have focused on the case of a projectile quark that radiates a gluon in response to medium-induced scattering. 
For an anti-quark projectile, the result is completely analogous and does not require separate discussion. For a gluon projectile, however, one finds
a larger set of distinct color configurations, as discussed already in section~\ref{sec2b} and shown in Figure~\ref{fig5}. Here we discuss this case explicitly for a gluon produced in the distant past. We first note that, identifying in Fig.~\ref{fig6} the projectile line with a gluon, three Feynman diagrams (a), (b) and (c) contribute to the radiation amplitude. To be specific, we label the color of the incoming gluon as  $d$. This gluon  splits into the most energetic `projectile'-fragment of color $b$ and a radiated gluon of color $a$. The color exchanged with the medium is labeled as $a_1$. In terms of  the generators $T_A$ of the adjoint representation, the corresponding contributions to the radiation amplitudes have the color structure:
\beq
{\cal M}_{(a)} \propto \left( T_A^a\, T_A^{a_1}  \right)_{bd}\, ,\qquad 
{\cal M}_{(b)} \propto \left( T_A^{a_1}\, T_A^a\right)_{bd}\, ,\qquad 
{\cal M}_{(c)} \propto \left( \left[T_A^a\, ,T_A^{a_1} \right]\right)_{bd}\, .
\label{eq3.14}
\eeq
Expressing the generators $T_A$ in terms of the structure constant of the Lie algebra, one can conveniently exploit relations like $(T^aT^{a_1})_{bd}=2\,{\rm Tr}\left([t^b,t^a][t^{a_1},t^d]\right)$ to identify the color channels involved. Due to the invariance of the trace for cyclic permutations only 6 distinct color configurations arise. There is a one-to-one correspondence between these 6 independent permutations and the six color differential diagrams for medium-induced gluon radiation shown in Figure~\ref{fig5}. One finds
\begin{alignat}{2}
	& {\rm ISR}(q) \leftrightarrow ba_1ad\, , &\qquad\qquad&{\rm ISR}(\bar{q}) \leftrightarrow bdaa_1\, ,\nonumber \\
	& {\rm FSR}(q) \leftrightarrow baa_1d\, , & &{\rm FSR}(\bar{q}) \leftrightarrow bda_1a \, ,\nonumber \\
	& {\rm R}q{\rm S}\bar{q} \leftrightarrow bada_1\, , & &{\rm R}\bar{q}{\rm S}q  \leftrightarrow ba_1da\, .
	\label{eq3.15}
\end{alignat}
In the calculation of the radiation cross section, interference terms between these six contributions correspond to non-planar diagrams and are subleading in the large-$N_c$ limit. In appendix~\ref{appa}, we provide explicit expressions for the amplitudes (\ref{eq3.14}) and their six distinct color-differential contributions (\ref{eq3.15}). From these expressions, one sees easily that the contributions $ {\rm R}q{\rm S}\bar{q}$ and $ {\rm R}\bar{q}{\rm S}q$ are suppressed by
a factor $O(x_g)$ compared to the other four terms. Up to $O(x_g)$ corrections (hence consistently with the approximation assumed in getting the color-inclusive GB spectrum), the terms  that describe initial or final state radiation off a q or $\bar{q}$-leg provide color-differential radiation cross sections of equal size. After summing over the final and averaging over the initial 
polarization of the gluons, each of these four contributions takes the form
\beq
\left.k^+\frac{dN}{dk^+ d\k}\right|_{[i]}=\frac{N_c}{4}\frac{\alpha_s}{\pi^2}\left\langle \frac{\q^2}{\k^2(\k-\q)^2}\right\rangle,
\eeq
with $i\!=\!{\rm FSR/ISR}(q/\bar{q})$. Summed together, these 4 leading contributions lead to the inclusive result:
\beq
k^+\frac{dN}{dk^+ d\k}=C_A\frac{\alpha_s}{\pi^2}\left\langle \frac{\q^2}{\k^2(\k-\q)^2}\right\rangle.
\eeq
In summary, these results show that at small $x_g$ and in the large $N_c$ limit, the $N=1$ radiation spectrum of a projectile gluon can be viewed as an incoherent superposition of the medium-induced radiation of a quark (with a silent anti-quark spectator), and of an anti-quark (with a silent quark as spectator). Modifications to this picture, arising from the qualitatively novel color configurations ($ {\rm R}q{\rm S}\bar{q}$ and $ {\rm R}\bar{q}{\rm S}q$), are suppressed in the soft limit by a factor $O(x_g^2)$.

\subsection{Color-differential medium-induced gluon radiation: the case $N=1$, $x_0^+= 0$}
\begin{figure}[!tp]
\begin{center}
\includegraphics[clip,width=.8\textwidth]{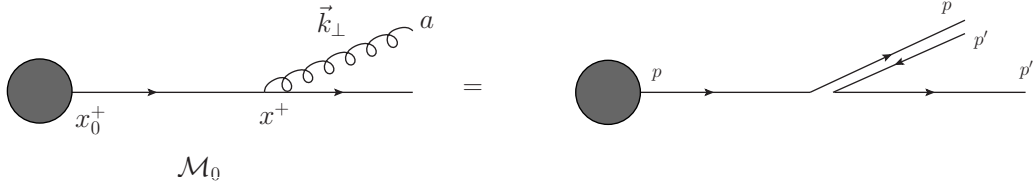}
\caption{The vacuum-radiation diagram: the dark blob denotes the hard process in which the high-momentum parton is produced (off-shell) inside the medium. In the right panel we keep track of the color flow.} 
\label{fig7}
\end{center}
\end{figure}

The gluon spectrum (\ref{gunionbertsch}) vanishes in the absence of a medium, consistently with the idea that a parton coming from the distant past is on-shell. The situation realized in a hadronic collision is different. There,  high-energy partons are produced around the time of the collision, at $x_0^+ = 0$ say, and they branch even in the absence of further in-medium interactions. Here, we calculate the medium-modification of this vacuum radiation. That is,  we consider the $N=1$ corrections in opacity to the `vacuum' splitting process shown in Figure~\ref{fig7}. The amplitude for this vacuum baseline is written as
%
%
\beq
i{\cal M}_0=(igt^a)[(2p_f\!+\!\kg)\!\cdot\!\eg]\frac{i}{(p_f+\kg)^2\!-\!M^2} J(p_f\!+\!\kg)e^{i(p_f+\kg)x_0}\, ,
\eeq
where the current $J(p_f\!+\!k_g)$ represents the hard production process. The resulting spectrum reads
\beq
k^+\frac{d\sigma^{\rm vac}}{dk^+d\kkg}=d\sigma^{\rm hard}\,C_R\,\frac{\alpha_s}{\pi^2}\K_0^2,
\eeq
with its characteristic collinear divergence in the massless limit and dead-cone effect suppressing the radiation of small-angle gluons in the case of emission off a massive quark, where $\K_0^2 = \kkg^2/ \left(\kkg^2 + x_g^2 M^2 \right)^2$. 
In general, the radiation amplitude can be expanded in the number of gluon exchanges with the medium,
\beq
i{\cal M}_{\rm rad}=i{\cal M}_0+i{\cal M}_1+i{\cal M}_2+\dots
\eeq
%
%
\begin{figure}[!tp]
\begin{center}
\includegraphics[clip,width=.7\textwidth]{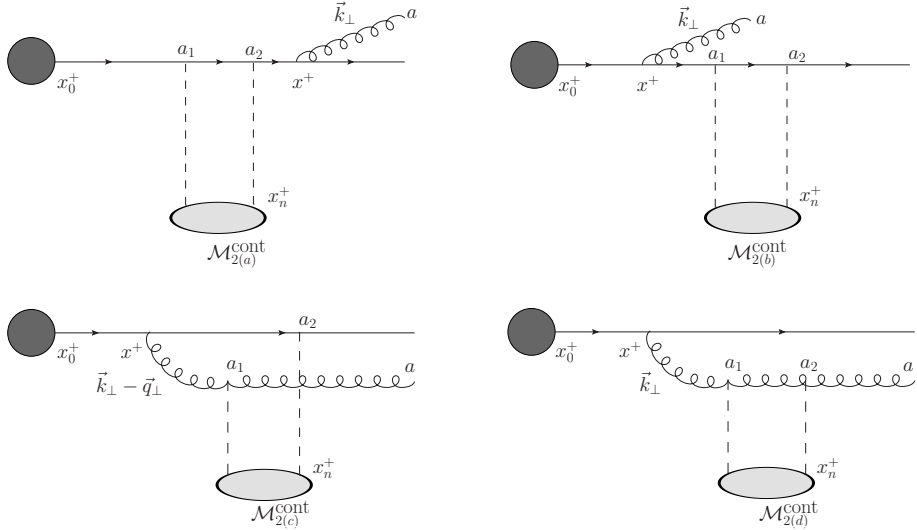}
\caption{The gluon-radiation amplitude arising from interacting twice with the same scattering center of the medium. In the $N=1$ opacity expansion, after the target average (over color and $\x_n$), the diagrams contribute to the single-gluon spectrum with an interference term decreasing the weight of the vacuum radiation.} 
\label{fig8}
\end{center}
\end{figure}
To perform a calculation that is complete to first order in opacity, the above expansion must include not only the complete set of one-gluon exchanges between partonic projectile and medium ${\cal M}_1$, but also a specific subset, which we label ${\cal M}_2^{\rm cont}$,  of two-gluon exchanges. 
Contributions to first order in opacity arise both from $\langle |{\cal M}_1|^2\rangle$ and ${\rm Re}\left\langle{\cal M}_2{\cal M}_0^*\right\rangle$.

The medium average $\langle \dots\rangle$ involves a color trace over the target and an integration over the transverse position of the scattering centers. For the two scattering centers that enter the term ${\rm Re}\left\langle{\cal M}_2{\cal M}_0^*\right\rangle$, these averages reduce to
\beq
{\rm Tr}\left(T^{a_1}_{(n)}T^{a_2}_{(n')}\right)\equiv\delta_{nn'}\delta^{a_1a_2}T_{F/A},\quad
\int d\x_ne^{-i(\q_1+\q_2)\cdot\x_n}=2\pi\delta(\q_1+\q_2)\,,
\eeq
where $T_F\!=\!(1/2)$ and $T_A = N_c$ for scattering off a quark or a gluon from the medium, respectively. 
As a consequence, after average over the target, the non-vanishing contributions to ${\rm Re}\left\langle{\cal M}_2{\cal M}_0^*\right\rangle$ arise from processes where the two gluons link to \emph{the same scattering center} in the medium and where neither color nor transverse momentum is transferred from medium to the fragmenting $q\to q + g$ system. It is usual to refer to these contributions as {\it contact terms} and they
are needed for a formulation that conserves probability, thus playing the role of virtual corrections. 
The contact terms contributing to ${\cal M}_2$ are shown in Figure~\ref{fig8}. 
To first order in opacity, one has
\beq
\left\langle|{\cal M}_0+{\cal M}_1+{\cal M}_2^{\rm cont}+...|^2\right\rangle=|{\cal M}_0|^2+\left\langle|{\cal M}_1|^2\right\rangle+2{\rm Re}\left\langle{\cal M}_2^{\rm cont}{\cal M}_0^*\right\rangle+...\, .
\label{firstopac}
\eeq
The radiation amplitude can be recast in a form more suitable for a color differential calculation: 
\beq
	{\cal M}_0+{\cal M}_1+{\cal M}_2^{\rm cont} =  {\cal M}^{a a_1} + {\cal M}^{a_1 a} + {\cal M}^a\,, 
	\label{firstopaccolor}
\eeq
where
\beq
{\cal M}_1= {\cal M}^{a a_1} + {\cal M}^{a_1 a}\quad{\rm and}\quad
{\cal M}_0+{\cal M}_2^{\rm cont}={\cal M}^a.
\eeq
In contrast to the left-hand side of (\ref{firstopaccolor}), interference terms between different contributions on the right hand side vanish to leading order in $N_c$. This allows us to derive, in the large-$N_c$ limit, the color-differential radiation cross sections by squaring the individual contributions from the various color channels.
More explicitly, the ${\cal M}^{a\, a_1} $ term reads
\begin{multline}
i{\cal M}^{aa_1}=-i\,g\,(t^at^{a_1})e^{ip^+x_0^-}\sum_{n=1}^{N}\theta(x_n^+\!-\!x_0^+)\int\frac{d\q}{(2\pi)^2}\,
{\cal A}(\q)\,T_{(n)}^{a_1}\,e^{-i\q\cdot\x_n}J(p^+)\\
\times\,e^{i\wob x_n^+}e^{i\frac{M^2}{2p_+}x_0^+}
\,2\,\eeg\!\cdot\!\left[\K_0-\left(1-e^{-i\wib (x_n^+-x_0^+)}\right)\K_1\right]\, .
\label{aa1}
\end{multline}
In this expression, the term proportional to $\K_0$ corresponds to a hard parton that emits a gluon after scattering; it is the term ${\cal M}_{1(a)}$ in Fig.~\ref{fig4}. 
On the other hand, the three-gluon vertex in the real emission amplitude ${\cal M}_{1(c)}$ in Fig.~\ref{fig4} gives rise to a term proportional to $\lbrack t^a,t^{a_1}\rbrack$, and the $t^a t^{a_1}$-part of this commutator provides the contribution $\propto \K_1$ in Eq.~(\ref{aa1}). The two $\K_1$ terms (with their different phase factors) arise from processes in which the propagator of the hard projectile parton is on or off-shell, respectively. In close analogy, one finds
\begin{multline}
i{\cal M}^{a_1a}=i\,g\,(t^{a_1}t^a)e^{ip^+x_0^-}\sum_{n=1}^{N}\theta(x_n^+\!-\!x_0^+)\int\frac{d\q}{(2\pi)^2}
\,{\cal A}(\q)\,T_{(n)}^{a_1}\,e^{-i\q\cdot\x_n}J(p^+)\\
\times\,e^{i\wob x_n^+}e^{i\frac{M^2}{2p_+}x_0^+}
\,2\,\eeg\!\cdot\!\left[\left(1\!-\!e^{-i\wob (x_n^+-x_0^+)}\right)\K_0-\left(1\!-\!e^{-i\wib (x_n^+-x_0^+)}\right)\K_1\right],
\label{codiff}
\end{multline}
where the term proportional to $\K_0$ corresponds to ${\cal M}_{1(b)}$ in Fig.~\ref{fig4} and the term proportional to $\K_1$ is the part of ${\cal M}_{1(c)}$ proportional to  $t^{a_1}t^a$. Again, the phase factors are different depending on whether the projectile from the hard event is on or off-shell.
In the color-differential amplitude (\ref{codiff}), the phase factors are written in terms of  the `transverse energies'
 \beq
 	\wob \equiv \frac{\kkg^2 + x^2 M^2}{2\, x_g\, p_+}\, ,\qquad \wib \equiv \frac{\left(\kkg- {\bf q}\right)^2 + x^2 M^2}{2\, x_g\, p_+}\, .
 \eeq 
In general, the inverse transverse energies $1/\wob$, $1/\wib$ act as formation times for medium-induced gluon emission. The color-averaged result for medium-induced gluon radiation to first order in opacity  is known to depend only on $1/\wib$ (we reproduce this result as a check in Eq.~(\ref{eq3.29}) below). 
 For the color-differential case studied here, we observe that the result depends on both formation times $1/\wob$ and $1/\wib$. In particular, after squaring the above amplitudes and averaging over the longitudinal position $x_n^+\in \left[x_0^+; x_0^+ + L^+ \right]$ for a medium of constant density, we find a {\it vacuum-like} contribution from the color-channel `$aa_1$' and a {\it medium-modified} one from the channel `$a_1a$':
\beqa
\langle|{\mathcal M}_1^{aa_1}|^2\rangle &\sim& \left\langle(\K_0-\K_1)^2+\K_1^2+2\K_1\!\cdot\!(\K_0-\K_1)\frac{\sin[\wib L^+]}{\wib L^+}\right\rangle\, ,
  \label{eq3.26}\\
\langle|{\mathcal M}_1^{a_1a}|^2\rangle &\sim& 2\left(1\!-\!\frac{\sin[\wob L^+]}{\wob L^+}\right)\K_0^2+2\left\langle\left(1\!-\!\frac{\sin[\wib L^+]}{\wib L^+}\right)\K_1^2\right\rangle
	\nonumber \\
	&& \qquad -
	\left\langle2\left(1\!-\!\frac{\sin[\wob L^+]}{\wob L^+}\!-\!\frac{\sin[\wib L^+]}{\wib L^+}\!+\!\frac{\sin[(\wib\!-\!\wob)L^+]}{(\wib\!-\!\wob) L^+} \right)\K_0\cdot\K_1\right\rangle .
	\label{eq3.27}
\eeqa
To first order in opacity, the calculation is completed by including the channel `$a$' in which  no color is exchanged with the medium. 
We have $\langle|{\mathcal M}_1^a|^2\rangle = |{\cal M}_0|^2 +2{\rm Re}\left\langle{\cal M}_2^{\rm cont}{\cal M}_0^*\right\rangle$, where $|{\cal M}_0|^2$ accounts for the vacuum branching and
\beqa
2{\rm Re}\left\langle{\cal M}_2{\cal M}_0^*\right\rangle &\sim&
-\K_0^2 - 2\left(1-\frac{\sin[\wob L^+]}{\wob L^+}\right)\K_0^2 \nonumber\\
 &&-2\left \langle \left(\frac{\sin[\wob L^+]}{\wob L^+}-\frac{\sin[(\wib-\wob) L^+]}{(\wib-\wob) L^+} \right)\K_0\!\cdot\!\K_1 \right\rangle\, .
  \label{eq3.28}
\eeqa
The sum $\langle|{\mathcal M}_1^{aa_1}|^2\rangle + \langle|{\mathcal M}_1^{a_1a}|^2\rangle + 2{\rm Re}\left\langle{\cal M}_2^{\rm cont}{\cal M}_0^*\right\rangle$  
combines to the color-averaged medium-induced  gluon radiation cross section
\beq
	 k^+\, \frac{dI^{\rm med}}{dk^+\, d\kg} = C_R\, \frac{\alpha_s}{\pi^2}\, \frac{L^+}{\lambda_{\rm el}^+} \, 
	 	\left\langle \left( \left(\K_0-\K_1\right)^2 - \K_0^2 + \K_1^2\right)   \left(1 -  \frac{\sin[\wib L^+]}{\wib L^+} \right)   \right\rangle\, .
		\label{eq3.29}
\eeq
This expression shows explicitly that the opacity expansion is an expansion in powers of $L^+/\lambda_{\rm el}^+$, where $\lambda_{\rm el}^+$ denotes the
elastic mean free path of a gluon and $R$ is the representation of the projectile. Here we are considering a quark, hence 
$C_R\!=\!C_F$.
%
\subsection{Dependence of color-differential $N=1$ gluon radiation on formation time(s)}
In the color-averaged cross section (\ref{eq3.29}), the phase factor 
\begin{equation}
  \left( 1 - \frac{\sin \left(\omega^-_1\, L^+ \right) }{\omega^-_1\, L^+} \right) =
          \left\{ \begin{array} 
           {r@{\quad  \hbox{for}\quad}l}
             0 & 1/\omega^-_1 \gg L^+ \, ,\\
             1 & 1/\omega^-_1 \ll L^+\, .
          \end{array} \right.
        	\label{eq3.30}
\end{equation} 
interpolates between the coherent and incoherent regimes.
Thus, the color-inclusive medium-induced spectrum in Eq.~(\ref{eq3.29}) vanishes if the gluon formation time prior to its rescattering $1/\omega^-_1$ is larger than the in-medium path length. This result matches the naive expectation that the medium can modify only the radiation of those gluons that are fully formed within its finite extension. Here, we discuss the different behavior of the various color channels: {\it vacuum-like} $(aa_1)$, {\it medium-modified} $(a_1a)$ and {not color-correlated} with the medium ($a$). They sum up to the inclusive result $dI\equiv dI^{\rm vac}+dI^{\rm med}$, where
\beq
k^+\frac{dI^{\rm vac}}{dk^+ d\kkg}\equiv\left.k^+\frac{dI^{\rm vac}}{dk^+ d\kkg}\right|_{a}=C_R\frac{\alpha_s}{\pi^2}\K_0^2
\eeq
and
\beq
k^+\frac{dI^{\rm med}}{dk^+ d\kkg}\equiv\left.k^+\frac{dI^{\rm med}}{dk^+ d\kkg}\right|_{aa_1}+
\left.k^+\frac{dI^{\rm med}}{dk^+ d\kkg}\right|_{a_1a}+\left.k^+\frac{dI^{\rm med}}{dk^+ d\kkg}\right|_{a}\,.
\label{eq3.31}
\eeq
For the color-differential contributions on the right hand side of this equation, the in-medium path length $L^+$ has to be compared with the two distinct formation times $1/\wob$ and $1/\wib$. To discuss this dependence, we focus on the four limiting cases in which either one or both formation times are either significantly larger or significantly smaller than $L^+$. In order to display more transparently the physical meaning of the various terms we will exploit the large-$N_c$ identities $C_A\!=\!2C_F$ for the color factors and $\lambda_q^+=2\lambda_g^+$ for the elastic mean free paths.
\begin{enumerate}
\item Totally incoherent case ($L^+\gg 1/\overline{\omega}_i$, with $i=0,1$)\\
In the limit $\overline{\omega}_iL^+ \to \infty$, the color-averaged medium-induced gluon radiation spectrum (\ref{eq3.29}) is proportional to $\left\langle  \left(\K_0-\K_1\right)^2 - \K_0^2 + \K_1^2 \right\rangle$. This is the incoherent superposition of: the usual Gunion-Bertsch spectrum for gluon emission $\propto \left\langle  \left(\K_0-\K_1\right)^2 \right\rangle$; a negative contribution to the vacuum radiation spectrum, $\propto - \left\langle  \K_0^2 \right\rangle$, that corrects the vacuum term for the probability that the radiated gluon interacts with the medium; and a  vacuum-radiation spectrum that is shifted in transverse momentum due to the rescattering of the radiated gluon in the medium  $\propto \left\langle  \K_1^2 \right\rangle$.
For the corresponding color differential contributions, we find from (\ref{eq3.26}), (\ref{eq3.27}) and (\ref{eq3.28})
\begin{subequations}
\begin{alignat}{2}
&\left.k^+\frac{dI^{\rm med}}{dk^+ d\kkg}\right|_{aa_1}&\,&\underset{\overline{\omega}_i L^+ \to\infty}{\sim}\frac{C_F}{2}\frac{\alpha_s}{\pi^2} \frac{L^+}{\lambda_{g}^+}
\left\langle(\K_0-\K_1)^2+\K_1^2\right\rangle\, , \label{subeq:aa1}\\
&\left.k^+\frac{dI^{\rm med}}{dk^+ d\kkg}\right|_{a_1a}&&\underset{\overline{\omega}_i L^+\to\infty}{\sim}
\frac{\alpha_s}{\pi^2}\!\left[\frac{L^+}{{\lambda_g^{+}}}\!\left(\frac{C_F}{2}\right)\left(\left\langle(\K_0\!-\!\K_1)^2\right\rangle\!+\!\left\langle\K_1^2\right\rangle\right)+\frac{L^+}{{\lambda_q^{\rm +}}}C_F\K_0^2\right]\, ,\label{subeq:a1a}\\
&\left.k^+\frac{dI^{\rm med}}{dk^+ d\kkg}\right|_{a}&&\underset{\overline{\omega}_iL^+ \to\infty}{\sim} \frac{C_F}{2}\frac{\alpha_s}{\pi^2}\frac{L^+}{\lambda_{g}^+}(-3\K_0^2)\label{subeq:a}\,.
\end{alignat}
\end{subequations}
After inspection of the color factors and of the mean-free-paths involved, the above terms admit a transparent physical interpretation. Eq.~(\ref{subeq:aa1}) is the sum of half of the GB spectrum $\langle(\K_0-\K_1)^2\rangle$ by an on-shell quark and half of the contribution of reshuffled vacuum radiation $\langle\K_1^2\rangle$ by an off-shell quark. Eq.~(\ref{subeq:a1a}), on top of this, gets an additional contribution from the vacuum radiation $\K_0^2$ by an off-shell quark which then suffers a further elastic scattering in the medium.
In this limit, the medium-modified color differential contribution `$a_1a$' is the largest. As a consequence,  the radiated gluon is decorrelated in color from the leading partonic fragment  in more than 
half of the medium-modified
parton branchings. This case is depicted for instance in Fig.~\ref{fig3}. 
	\item Totally coherent case ($1/\overline{\omega}_i\gg L^+$)\\
	For very large gluon formation times, the gluon is produced far outside the medium. At the level of the color-inclusive result the medium is not a source of an enhanced rate of parton branching: the color-averaged medium-induced spectrum in Eq.~(\ref{eq3.29}) vanishes.
However, with the probability that an elastic interaction occurs, the fragmenting projectile is color connected to the medium (`$aa_1$' channel) rather than to the hard production process (`$a$' channel):
\begin{subequations}
\begin{alignat}{2}
&\left.k^+\frac{dI^{\rm med}}{dk^+ d\kkg}\right|_{aa_1}&\,&\underset{\overline{\omega}_iL^+\to 0}{\sim}\frac{L^+}{\lambda_{q}^+}C_F\frac{\alpha_s}{\pi^2}\K_0^2\, ,\label{subeq2:aa1}\\
&\left.k^+\frac{dI^{\rm med}}{dk^+ d\kkg}\right|_{a_1a}&&\underset{\overline{\omega}_iL^+\to 0}{\sim}0\, ,\label{subeq2:a1a}\\
&\left.k^+\frac{dI^{\rm med}}{dk^+ d\kkg}\right|_{a}&&\underset{\overline{\omega}_iL^+\to 0}{\sim}-\frac{L^+}{\lambda_{q}^+}C_F\frac{\alpha_s}{\pi^2}\K_0^2\,.\label{subeq2:a}
\end{alignat}
\end{subequations}
The rearrangement between the `$aa_1$' and `$a$' channels reflects the fact that color is exchanged between medium and projectile even if the color averaged spectrum (\ref{eq3.29}) remains unchanged, i.e. even if there is no overall enhanced probability of gluon radiation induced by the medium. Notice, however, that in both channels 
the gluon is color correlated with the highest-$p_T$ fragment.
	\item $1/\wib\gg L^+\gg 1/\wob $\\
	In this kinematic range,  the color-inclusive medium-induced radiation vanishes. Remarkably, however, the color connection amongst the most energetic fragments is still medium-modified and two-thirds of the gluons radiated due to medium effects (and compensated by a corresponding depletion in the vacuum spectrum) are carried by the `$a_1a$' contribution and are thus decorrelated in color from the most energetic fragment
\begin{subequations}
\begin{alignat}{2}
&\left.k^+\frac{dI^{\rm med}}{dk^+ d\kkg}\right|_{aa_1}&\,&\overset{\wob L^+\to \infty}{\underset{\wib L^+\to 0}{\sim}}\frac{L^+}{\lambda_{q}^+}C_F\frac{\alpha_s}{\pi^2}\K_0^2\, ,\label{subeq3:aa1}\\
&\left.k^+\frac{dI^{\rm med}}{dk^+ d\kkg}\right|_{a_1a}&&\overset{\wob L^+\to \infty}{\underset{\wib L^+\to 0}{\sim}}2\frac{L^+}{\lambda_{q}^+}C_F\frac{\alpha_s}{\pi^2}\K_0^2\, ,\label{subeq3:a1a}\\
&\left.k^+\frac{dI^{\rm med}}{dk^+ d\kkg}\right|_{a}&&\overset{\wob L^+\to \infty}{\underset{\wib L^+\to 0}{\sim}}- 3\frac{L^+}{\lambda_{q}^+}C_F\frac{\alpha_s}{\pi^2}\K_0^2\, .\label{subeq3:a} 
\end{alignat}
\end{subequations}
The factor 2 difference between the `$aa_1$' and `$a_1a$' channels reflects the fact that Eq.~(\ref{subeq3:aa1}) gets contribution only from processes in which the propagator of the hard quark is on-shell prior to scattering and radiation; for contributions to Eq.~(\ref{subeq3:a1a}) this propagator can be either on- or off-shell.
	
	\item $1/\wob\gg L^+\gg 1/\wib $\\
Also in this case, the medium-modified color channel `$a_1a$' -- in which the gluon decorrelates in color from the leading parton -- tends to be the most likely one for parton splitting
\begin{subequations}
\begin{alignat}{2}
&\left.k^+\frac{dI^{\rm med}}{dk^+ d\kkg}\right|_{aa_1}&\,&\overset{\wob L^+\to 0}{\underset{\wib L^+\to \infty}{\sim}}\frac{L^+}{\lambda_{g}^+}\frac{C_F}{2}\frac{\alpha_s}{\pi^2}
	\left\langle\left(\K_0-\K_1\right)^2 + \K_1^2 \right\rangle\, ,\\
&\left.k^+\frac{dI^{\rm med}}{dk^+ d\kkg}\right|_{a_1a}&&\overset{\wob L^+\to 0}{\underset{\wib L^+\to \infty}{\sim}}\frac{L^+}{\lambda_{g}^+}C_F\frac{\alpha_s}{\pi^2}\langle\K_1^2\rangle\, ,\\
&\left.k^+\frac{dI^{\rm med}}{dk^+ d\kkg}\right|_{a}&&\overset{\wob L^+\to 0}{\underset{\wib L^+\to \infty}{\sim}} \frac{L^+}{\lambda_{g}^+}\frac{C_F}{2}\frac{\alpha_s}{\pi^2}
\left\langle - \K_0^2 - 2 \K_0\cdot\K_1\right\rangle\, .
\end{alignat}
\end{subequations}
As an aside, we note that there are kinematic arguments for why this limiting case may be less relevant for a medium-modified parton shower\footnote{We 
recall that the gluon in the final state has transverse momentum ${\kkg}$ and therefore a formation time $1/\wob$. The term $1/\wib$ can thus be viewed as the
formation time of a gluon that did not yet undergo scattering with the medium. Since gluon emission is dominated by collinear branching, the initial gluon transverse momentum $\kkg-\q$ will be small in most branching processes, and a further momentum transfer from the medium is more likely to increase the gluon's transverse momentum than to reduce it. So, on qualitative grounds, the ordering $\vert\kkg-\q\vert < \vert \kkg\vert$ (and equivalently $1/\wob < 1/\wib$) is more likely to occur than the opposite one.}.
\end{enumerate}
In summary, whenever medium-induced gluon radiation is sizable, the qualitatively novel medium-modified color connection `$a_1a$' of the projectile with the medium arises in more than half of the processes. Moreover, even in cases in which the color-inclusive induced spectrum (\ref{eq3.29}) is negligible, the presence of the medium can act as a source of color-decorrelation between the radiated gluon and the leading parton. 


\section{Higher orders in opacity: the $N=2$ case}
\label{sec4}
For a parton shower that develops in the vacuum, the two daughters of a $q\to qg$ or $g \to gg$ branching process are always color connected. In contrast, in the $N=1$ opacity expansion carried out in the previous section, one finds that the interaction with the medium during a branching process decoheres the daughters in exactly half
of the cases for a projectile parton produced in the distant past ($x_0^+= - \infty$), and in more than 
half of all cases whenever there is a medium-induced gluon radiation from a parton produced in the medium ($x_0^+= 0$ and $1/\wib\ll L^+$). Here, we discuss how these findings change once more than one gluon exchange between the partonic projectile and the medium is taken into account. 

In color-inclusive calculations, parton energy loss is largely determined by the average squared transverse momentum per unit path-length irrespectively of whether the momentum is transferred in one or several gluon exchanges. 
The fraction of radiated gluons that are color decorrelated from the most energetic projectile fragment may, however, be expected to increase with the number of interactions with the medium. Here, we support this expectation by performing a color-differential calculation of the $N=2$ opacity contribution to gluon radiation by a parton produced in the distant past.

As usual, the starting point is the expansion of the radiation amplitude in the number of gluons exchanged with the medium:
\beq
{\cal M}^{\rm rad}={\cal M}_0+{\cal M}_1+{\cal M}_2^{\rm dir}+{\cal M}_2^{\rm virt}+{\cal M}_3^{\rm dir}+{\cal M}_3^{\rm virt}+...\,.
\label{eq4.10}
\eeq
As was already the case at $N=1$, the above amplitude involves `virtual' contact terms in which two gluons are exchanged between the projectile and a single scattering center in the medium, without net exchange neither of color nor of transverse momentum. Up to second order in opacity -- when the projectile arrives on-shell from the far past, so that ${\cal M}_0=0$ -- the radiation spectrum receives contributions from
\beq
|{\cal M}^{\rm rad}|^2=|{\cal M}_1|^2+|{\cal M}_2^{\rm dir}|^2+|{\cal M}_2^{\rm virt}|^2+2{\rm Re}\,{\cal M}_3^{\rm virt}{\cal M}_1^*+{\mathcal O}(L/\lambda_{\rm el})^3\, .
\label{eq4.11}
\eeq
Here, a contribution to zeroth order in opacity is absent for a projectile coming from the distant past, since a radiated gluon cannot be emitted on-shell without interaction with the medium. For the same reason, interference terms $2{\rm Re}\,{\cal M}_2^{\rm virt}{\cal M}_0^*$ vanish in $|{\cal M}^{\rm rad}|^2$. However, the contribution $|{\cal M}_2^{\rm virt}|^2$ does not vanish since a color-neutral two-gluon exchange between the projectile and the medium can -- in general -- still transfer longitudinal momentum, allowing the radiation of a gluon. Thus the amplitude for a color-neutral two-gluon exchange with no transfer of momentum $\q$ with one scattering center, paired with a corresponding term in the complex conjugate amplitude, leads to a finite contribution.
\subsection{Direct contributions to $N=2$}
\label{sec:N2direct}
\begin{figure}[!tp]
\begin{center}
\includegraphics[clip,width=1.0\textwidth]{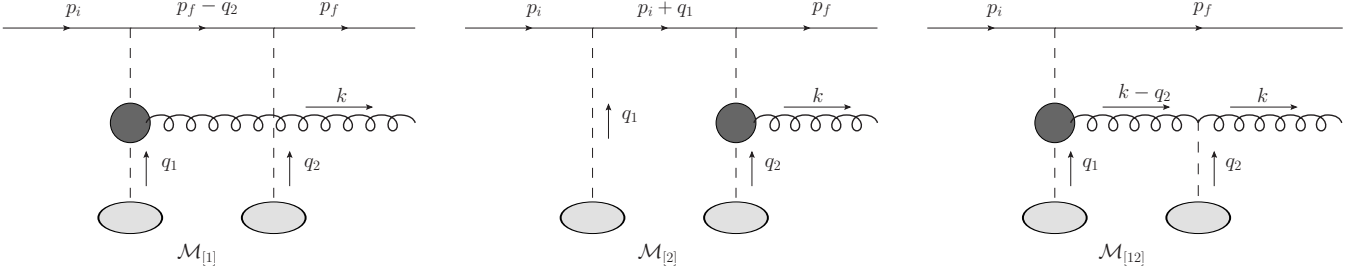}
\caption{The three gluon-radiation amplitudes involved in the $N=2$ scattering case. }
\label{fig9}
\end{center}
\end{figure}
To second order in opacity, the contributions involving two finite momentum transfers are depicted in Figure~\ref{fig9}. Here and in what follows, we represent effective non-local gluon emission vertices as dark blobs. They denote the combination of the three radiation amplitudes in Fig.~\ref{fig6} and correspond to the Lipatov vertices in BFKL calculations.
Their contribution to the radiation of a gluon of momentum $\kkg$ after momentum transfer ${\q}$ from the medium can be expressed in terms of the currents 
\beq
\J(\kg,q)\equiv\frac{\kkg}{\kkg^2+\xg^2M^2}-\frac{\kkg\!-\!\q}{(\kkg-\q)^2+\xg^2M^2}\, .
\eeq
We also introduce the notational shortcuts
\beq
\J_1\equiv\J(\kg,q_1),\quad \J_2\equiv\J(\kg,q_2)\quad{\rm and}\quad \J_{12}\equiv\J(\kg-q_2,q_1).
\eeq
Employing the above currents it is possible to express the $N=2$ direct contribution in the convenient form suggested by Fig.~\ref{fig9}
\beq
{\cal M}_2^{\rm dir}={\mathcal M}_{[1]}+{\mathcal M}_{[2]}+{\mathcal M}_{[12]},
\eeq
with
\begin{subequations}
\begin{align}
i{\mathcal M}_{[1]}\sim & g\, t^{a_2}[t^a,t^{a_1}]\,2\,\eeg\!\cdot\!\J_1\, e^{i\,\wob x_1^+}\quad\quad\quad\quad\quad\quad\label{subeq:1}\\
i{\mathcal M}_{[2]}\sim & g\, [t^a,t^{a_2}]t^{a_1}\,2\,\eeg\!\cdot\!\J_2\,e^{i\,\wob x_2^+}\quad\quad\quad\quad\quad\quad\label{subeq:2}\\
i{\mathcal M}_{[12]}\sim & g\, [[t^a,t^{a_2}],t^{a_1}]\,2\,\eeg\!\cdot\!\J_{12}\,e^{i\,\wiib x_1^+}e^{i\,(\wob-\wiib) x_2^+}.\label{subeq:12}
\end{align}
\end{subequations}

\begin{figure}[!tp]
\begin{center}
\includegraphics[clip,width=1.0\textwidth]{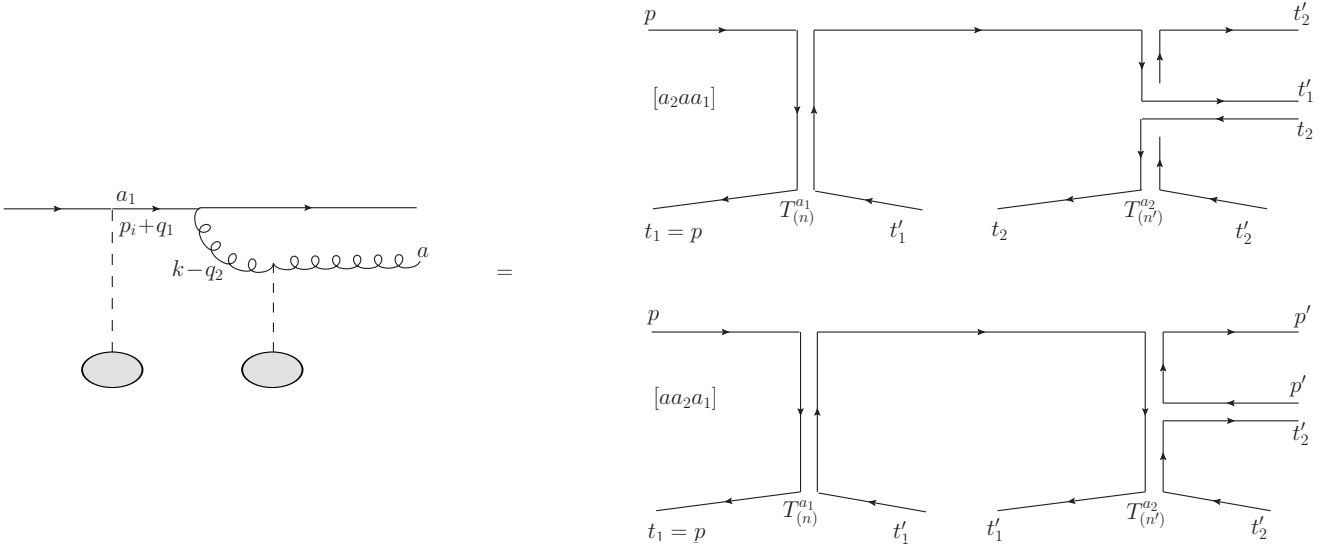}
\caption{An example of a Feynman diagram contributing to gluon radiation at $N=2$ order in opacity, with the corresponding color-flows associated to it. Two different processes are accounted for: the radiation of a gluon from the second scattering center, after the incoming on-shell quark already suffered an elastic collision (`$aa_2a_1$' contribution to ${\cal M}_{[2]}$) and the radiation of a on-shell gluon from the first scattering center, which then suffers a further elastic collision (`$a_2aa_1$' contribution to  ${\cal M}_{[12]}$).} 
\label{fig:N2gluon3}
\end{center}
\end{figure}
It follows from the color structure of the above amplitudes that the $N=2$ direct contribution can be organized into five distinct color channels 
\begin{equation}
	\left[a_2\, a\, a_1\right]\, ,\quad \left[a_2\, a_1\, a\right]\, ,\quad \left[a\, a_2\, a_1\right]\, ,\quad \left[a_1\, a\, a_2\right]\, ,\quad \left[a_1\, a_2\, a\right]\, .
\end{equation}
An example is shown in Fig.~\ref{fig:N2gluon3}.
The {\it vacuum-like} contribution, in which the color flows in the large $N_c$-limit from the projectile quark to the gluon, corresponds to the term $\left[a\, a_2\, a_1\right]$. In the four other color channels, color flows from the leading quark projectile directly to the medium without passing through the gluon, i.e., the gluon is color decohered from the quark. 
To leading order in $N_c$, these five color channels do not interfere.
From Eqs.~(\ref{subeq:1}-\ref{subeq:12}) one readily obtains the radiation spectrum in the different color channel (interference terms being suppressed by $1/N_c^2$ factors).
Transverse momentum kicks from the medium are weighted by the corresponding elastic cross-section, employing the shorthand notation
\beq
\langle\langle...\rangle\rangle_{\q_1,\q_2}\equiv\int\!\! d\q_1\left(\frac{1}{\sigma^{\rm el}}\frac{d\sigma^{\rm el}}{d\q_1}\right)\int\!\! d\q_2\left(\frac{1}{\sigma^{\rm el}}\frac{d\sigma^{\rm el}}{d\q_2}\right)\, .
\eeq
Furthermore, we introduce the current $|J_{-\infty}|^2\!\equiv\!(2p^+)^2$ describing the parton coming from the far past. The vacuum-like contribution takes the form (after averaging over the longitudinal location of the scattering centers)
\beq
\langle|{\mathcal M}_2^{aa_2a_1}|^2\rangle
=4g^2\, C_F\,\left(\frac{L^+}{\lambda_{q}^+}\right)^2\,|J_{-\infty}|^2\,\left\langle\left\langle\frac{1}{2}|\J_2|^2+\frac{1}{2}|\J_{12}|^2+2\frac{1-\cos(\wiib L^+)}{(\wiib L^+)^2} \J_2\!\cdot\!\J_{12}\right\rangle\right\rangle_{\q_1,\q_2}
\eeq
Amongst the four other color channels, three  do not carry phase interference terms,
\beqa
\langle|{\mathcal M}_2^{a_2a_1a}|^2\rangle
&=&4g^2\,C_F\left(\frac{L^+}{\lambda_{q}^+}\right)^2\,|J_{-\infty}|^2 \,\left\langle\left\langle\frac{1}{2}|\J_1|^2\right\rangle\right\rangle_{\q_1,\q_2}\, ,\nonumber\\
\langle|{\mathcal M}_2^{a_1aa_2}|^2\rangle&=&
\langle|{\mathcal M}_2^{a_1a_2a}|^2\rangle
= 4g^2\, C_F\,\left(\frac{L^+}{\lambda_{q}^+}\right)^2\,|J_{-\infty}|^2\,\left\langle\left\langle\frac{1}{2}|\J_{12}|^2\right\rangle\right\rangle_{\q_1,\q_2}\, ,
\eeqa
while the fourth one has a slightly more complicated structure
\begin{multline}
\langle|{\mathcal M}_2^{a_2aa_1}|^2\rangle
=4g^2\,C_F\,\left(\frac{L^+}{\lambda_{q}^+}\right)^2\,|J_{-\infty}|^2\,
\left\langle\left\langle
\frac{1}{2}|\J_1|^2+\frac{1}{2}|\J_2|^2+\frac{1}{2}|\J_{12}|^2
-2\frac{1-\cos(\wob L^+)}{(\wob L^+)^2} \J_1\!\cdot\!\J_{2}\right.\right.\\
\left.\left.-2\frac{1-\cos(\woiib L^+)}{(\woiib L^+)^2} \J_1\!\cdot\!\J_{12}
+2\frac{1-\cos(\wiib L^+)}{(\wiib L^+)^2} \J_2\!\cdot\!\J_{12}
\right\rangle\right\rangle_{\q_1,\q_2}\, .
\end{multline}
The sum of the above five terms yields the $N=2$ \emph{direct} contribution to the spectrum:
\begin{eqnarray}
\langle|{\cal M}_2^{\rm dir}|^2\rangle&=&4g^2\,C_F\,\left(\frac{L^+}{\lambda_{\rm q}^+}\right)^2\,|J_{-\infty}|^2
\left\langle\left\langle|\J_1|^2+|\J_2|^2+2|\J_{12}|^2 -2\frac{1-\cos(\wob L^+)}{(\wob L^+)^2} \J_1\!\cdot\!\J_{2} \right.\right.  \nonumber\\
&&\left.\left.-2\frac{1-\cos(\woiib L^+)}{(\woiib L^+)^2} \J_1\!\cdot\!\J_{12}
+4\frac{1-\cos(\wiib L^+)}{(\wiib L^+)^2} \J_2\!\cdot\!\J_{12}\right\rangle\right\rangle_{\q_1,\q_2}\, .
\label{eq4.7}
\end{eqnarray}
From this and the corresponding color-differential expressions, one then checks 
easily that, in the totally incoherent limit, the vacuum-like contribution
\beq
\langle|{\mathcal M}_2^{aa_2a_1}|^2\rangle\underset{\overline{\omega}_iL^+\to\infty}{\sim} \frac{1}{2}\left\langle|\J_2|^2+|\J_{12}|^2\right\rangle
\label{eq4.8}
\eeq
is always more than a factor 2 smaller than the sum over all five color channels
\beq
\langle|{\mathcal M}^{\rm dir}_2|^2\rangle\underset{\overline{\omega}_iL^+\to\infty}{\sim}\left\langle|\J_1|^2+|\J_2|^2+2|\J_{12}|^2\right\rangle\,.
\label{eq4.9}
\eeq
This is consistent with the naive expectation that a vacuum-like fragmentation pattern is less likely to survive in higher orders in opacity. 
\subsection{Virtual contributions to $N=2$}
The $N=2$ opacity calculation is completed by the computation of the relevant contact terms. In the case at hand, a projectile arriving on-shell from the far past, these correspond to the terms $\langle|{\cal M}_2^{\rm virt}|^2\rangle$ and $2{\rm Re}\langle{\cal M}_3^{\rm virt}{\cal M}_1^*\rangle$ in Eq.~(\ref{eq4.11}). From the color point of view they contribute, respectively,  to the `$a$' and `$aa_1$'/`$a_1a$' channels:
\beqa
\langle|{\cal M}_2^{\rm virt}|^2\rangle&\equiv&\langle|{\cal M}_2^{\rm virt}|^2\rangle_a\, ,\nonumber\\
2{\rm Re}\langle{\cal M}_3^{\rm virt}{\cal M}_1^*\rangle&\equiv&2{\rm Re}\langle{\cal M}_3^{\rm virt}{\cal M}_1^*\rangle_{aa_1}+2{\rm Re}\langle{\cal M}_3^{\rm virt}{\cal M}_1^*\rangle_{a_1a}.
\eeqa
The term $\langle|{\cal M}_2^{\rm virt}|^2\rangle_a$ accounts for processes in which neither color nor transverse momentum is exchanged between projectile and medium, but where the longitudinal (`-' in light-cone coordinates) momentum transferred by the medium is sufficient to open the possibility of gluon radiation for a quark initially on-shell. This is a novel contribution, absent at N=1 opacity.
On the other hand, the terms $2{\rm Re}\langle{\cal M}_3^{\rm virt}{\cal M}_1^*\rangle_{aa_1/a_1a}$ are virtual corrections which combine with the direct amplitudes $\langle|{\cal M}_1^{aa_1}|^2\rangle$ and $\langle|{\cal M}_1^{a_1a}|^2\rangle$, and that decrease the weight of the corresponding color channels.

The calculation of $\langle|{\cal M}_2^{\rm virt}|^2\rangle_a$ requires the evaluation of the same contact terms, shown in Fig.~\ref{fig8},  as in the $N=1$ case for a quark produced at $x_0^+=0$. Here, however,  only the terms referring to a quark arriving on-shell are retained
\begin{equation}
\langle|{\cal M}_2^{\rm virt}|^2\rangle_{a}= 4g^2\,C_F\,\left(\frac{L^+}{\lambda_{q}^+}\right)^2\,|J_{-\infty}|^2\,\left\langle\left\langle 2\frac{1-\cos(\wob L^+)}{(\wob L^+)^2} \J_1\!\cdot\!\J_{2}\right\rangle\right\rangle_{\q_1,\q_2}.
\label{eq:N2a}
\end{equation}
In the color inclusive spectrum, this contribution cancels the $\wob$-dependent terms in (\ref{eq4.7}).
We further note that most `virtual' corrections in the opacity expansion can be regarded as probability conserving terms that subtract yield from terms of lower order in opacity. 
In contrast, the contribution (\ref{eq:N2a}) is always positive, and it opens a channel `$a$' absent at lower order in opacity. 

\begin{figure}[!htp]
\begin{center}
\includegraphics[clip,width=0.6\textwidth]{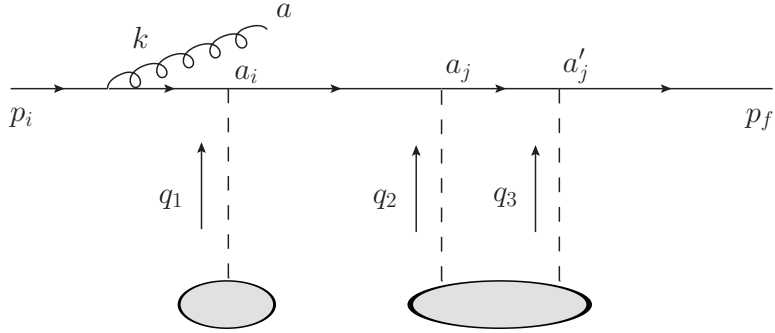}
\caption{An example of the contact diagrams contributing to ${\cal M}_3^{\rm virt}$.}
\label{fig:N3contact}
\end{center}
\end{figure}
The calculation of the term $2{\rm Re}\langle{\cal M}_3^{\rm virt}{\cal M}_1^*\rangle$ involves the evaluation of diagrams like the one in Fig.~\ref{fig:N3contact}. Details of the calculation are given in Appendix \ref{app:n2}. One gets for the two channels 
\begin{multline}
2{\rm Re}\langle{\cal M}_3^{\rm virt}{\cal M}_1^*\rangle_{aa_1}=4g^2\,C_F\,\left(\frac{L^+}{\lambda_{\rm q}^+}\right)^2\,|J_{-\infty}|^2\,\left\langle\left\langle-\frac{3}{2}|\J_1|^2-\frac{1}{2}|\J_2|^2+2\frac{1-\cos(\wob L^+)}{(\wob L^+)^2} \J_1\!\cdot\!\J_{2}\right.\right. \\
\left.\left.+2\frac{1-\cos(\woiib L^+)}{(\woiib L^+)^2} \J_1\!\cdot\!\J_{12}+2\frac{1-\cos(\wiib L^+)}{(\wiib L^+)^2} \J_2\!\cdot\!\J_{12} \right\rangle\right\rangle_{\q_1,\q_2}\, ,
\label{eq:N=2aa1}
\end{multline}
and
\begin{multline}
2{\rm Re}\langle{\cal M}_3^{\rm virt}{\cal M}_1^*\rangle_{a_1a}
=4g^2\,C_F\,\left(\frac{L^+}{\lambda_{q}^+}\right)^2\,|J_{-\infty}|^2\times\\
\left\langle\left\langle-\frac{3}{2}|\J_1|^2-\frac{1}{2}|\J_2|^2-2\frac{1-\cos(\wob L^+)}{(\wob L^+)^2} \J_1\!\cdot\!\J_{2}+2\frac{1-\cos(\wiib L^+)}{(\wiib L^+)^2} \J_2\!\cdot\!\J_{12} \right\rangle\right\rangle_{\q_1,\q_2}\, .
\label{eq:N=2a1a}
\end{multline}
%
\subsection{Gluon radiation up to order $N\!=\!2$ in opacity}
We can now write the medium-induced radiation spectrum up to order $N=2$ in opacity. 
It can be written as $dI=dI^{N=1}+dI^{N=2}$, where $dI^{N=1}$ is the lowest-order ($N=1$ opacity) result:
\beq
k^+\frac{dI^{N=1}}{dk^+ d\k}=C_A\frac{\alpha_s}{\pi^2}\left(\frac{L^+}{\lambda_{\rm q}^+}\right)\left\langle|\J_1|^2\right\rangle_{\q_1}\, ,
\eeq
while $dI^{N=2}$ arises from the sum $\langle|{\cal M}_2^{\rm dir}|^2\rangle+\langle|{\cal M}_2^{\rm virt}|^2\rangle+2{\rm Re}\langle{\cal M}_3^{\rm virt}{\cal M}_1^*\rangle$. 
In the large-$N_c$ limit when $C_A=2C_F$ for the color charges and $\lambda_q=2\lambda_g$ for the mean free paths, one finds
\beq
k^+\frac{dI^{N=2}}{dk^+ d\k}=C_A\frac{\alpha_s}{\pi^2}\left(\frac{L^+}{\lambda_{q}^+}\right)^2\left\langle\left\langle|\J_{12}|^2-|\J_1|^2+2\frac{1-\cos(\wiib L^+)}{(\wiib L^+)^2}2\J_2\!\cdot\!\J_{12}\right\rangle\right\rangle_{\q_1,\q_2},
\eeq
which coincides with the well known result quoted for instance in Ref.~\cite{Zapp:2008gi}\footnote{Note that in~\cite{Zapp:2008gi} Minkowski coordinates are employed and the mean-free-path $\lambda_q$ is replaced by $2\lambda_g$.}.

While the above color-inclusive result depends on a single formation time $1/\wiib$,  the color-differential contributions to second order in opacity depend already on three different formation times and have a somewhat more complicated structure.
Let us now express the radiation spectrum in a color-differential way.

The channels `$aa_1$' and `$a_1a$', already present at order $N=1$, receive in the $N=2$ calculation a correction from the virtual terms 
\begin{subequations}
\begin{align}
dI_{aa_1}&=dI_{aa_1}^{N=1}+dI_{aa_1}^{N=2}\\
dI_{a_1a}&=dI_{a_1a}^{N=1}+dI_{a_1a}^{N=2}.
\end{align}
\end{subequations}
In the above $dI_{aa_1}^{N=1}=dI_{a_1a}^{N=1}=(1/2)dI^{N=1}$, while the $N=2$ corrections can be obtained from Eqs.~(\ref{eq:N=2aa1}) and (\ref{eq:N=2a1a}) and read:
\begin{multline}
\left.k^+\frac{dI^{N=2}}{dk^+ d\k}\right|_{\rm aa_1}=C_F\frac{\alpha_s}{\pi^2}\left(\frac{L^+}{\lambda_{q}^+}\right)^2\left\langle\left\langle-\frac{3}{2}|\J_1|^2-\frac{1}{2}|\J_2|^2+2\frac{1-\cos(\wob L^+)}{(\wob L^+)^2} \J_1\!\cdot\!\J_{2}\right.\right.\\
\left.\left.+2\frac{1-\cos(\woiib L^+)}{(\woiib L^+)^2} \J_1\!\cdot\!\J_{12}+2\frac{1-\cos(\wiib L^+)}{(\wiib L^+)^2} \J_2\!\cdot\!\J_{12} \right\rangle\right\rangle_{\q_1,\q_2},
\end{multline}
\begin{multline}
\left.k^+\frac{dI^{N=2}}{dk^+ d\k}\right|_{\rm a_1a}=C_F\frac{\alpha_s}{\pi^2}\left(\frac{L^+}{\lambda_{q}^+}\right)^2\left\langle\left\langle-\frac{3}{2}|\J_1|^2-\frac{1}{2}|\J_2|^2-2\frac{1-\cos(\wob L^+)}{(\wob L^+)^2} \J_1\!\cdot\!\J_{2}\right.\right.\\
\left.\left.+2\frac{1-\cos(\wiib L^+)}{(\wiib L^+)^2} \J_2\!\cdot\!\J_{12} \right\rangle\right\rangle_{\q_1,\q_2}.
\end{multline}

Furthermore the $N=2$ calculation open new channels absent at lower order in opacity.
The spectrum in the `$a$'-channel, involving no color exchange, follows from Eq.~(\ref{eq:N2a}):
\beq
\left.k^+\frac{dI^{N=2}}{dk^+ d\k}\right|_{\rm a}=C_F\frac{\alpha_s}{\pi^2}\left(\frac{L^+}{\lambda_{q}^+}\right)^2\left\langle\left\langle 2\frac{1-\cos(\wob L^+)}{(\wob L^+)^2} \J_1\!\cdot\!\J_{2}\right\rangle\right\rangle_{\q_1,\q_2}.
\eeq
Concerning the channels arising from ${\cal M}_2^{\rm dir}$ one has, for instance,  the channel with {\it vacuum-like} color connection of the leading fragment with the radiated gluon,
\beq
\left.k^+\frac{dI^{N=2}}{dk^+ d\k}\right|_{\rm aa_2a_1}=C_F\frac{\alpha_s}{\pi^2}\left(\frac{L^+}{\lambda_{q}^+}\right)^2\left\langle\left\langle\frac{1}{2}|\J_2|^2+\frac{1}{2}|\J_{12}|^2+2\frac{1-\cos(\wiib L^+)}{(\wiib L^+)^2} \J_2\!\cdot\!\J_{12}\right\rangle\right\rangle_{\q_1,\q_2}\, .
\eeq
The spectrum in the other channels can be easily obtained from the corresponding squared amplitudes quoted in Sec.~\ref{sec:N2direct}.

Here, we end by considering the totally incoherent limit in which $L^+$ is much larger than all formation times in the problem. This is the dominant phase space region for gluon radiation in the sense that gluon radiation is not suppressed by destructive quantum interference. One checks easily that, in this totally incoherent limit, the totally virtual term $\langle|{\cal M}_2^{\rm virt}|^2\rangle$ (the one contributing to the `$a$'-channel) vanishes. The vacuum-like `$aa_1$' and medium-modified `$a_1a$' probability conserving factors $2{\rm Re}\langle{\cal M}_3^{\rm virt}{\cal M}_1^*\rangle$ have equal negative weight $\propto - 3 |J_1|^2/2 - |J_2|^2/2$. However, for the terms with two finite transverse momentum transfers from the medium, $\langle|{\cal M}_2^{\rm dir}|^2\rangle$, the contributions with medium-modified color-flow (the ones in which the gluon is color-decohered) have a weight $\propto \left( 2 |J_1|^2 + |J_2|^2 + 3 |J_{12}|^2\right)$ that is larger than the vacuum-like one $\propto \left(  |J_2|^2 +  |J_{12}|^2\right)$. This illustrates with an explicit calculation the general expectation that the weight of vacuum-like contributions to the total radiation spectrum decreases as higher orders in opacity are accounted for.
\subsection{General features of a color-differential analysis of medium-induced gluon radiation}
\label{sec:features}
We now summarize what we learned from the technical analysis performed in the previous sections, in which we calculated medium-induced gluon radiation in a spatially extended QCD medium within an opacity expansion. The color-inclusive result is consistent with previously known expressions.
In addition, our calculation provides for the first time color-differential information about medium-induced gluon radiation. 
We distinguish in particular {\it vacuum-like} contributions (in which the leading final state parton is color correlated with the radiated gluon)
from {\it medium-modified} contributions (in which the radiated gluon is decohered in color from the projectile)~\footnote{More precisely,
we call the large-$N_c$ limit of a radiation-amplitude vacuum-like if and exactly if the quark line of the leading final state quark (or one of the
two quark and anti-quark lines of the most energetic final state gluon) connect to the radiated gluon without passing through components
of the medium.}.
Up to 2nd order in opacity, we found three vacuum-like contributions 
\begin{equation}
	\left[a\right]\, ,\qquad \left[a\, a_1\right]\, ,\qquad \left[ a\, a_2\, a_1\right]\, .
\end{equation}
Here, the notation $\left[a\, ...\right]$ indicates contributions to the gluon radiation amplitude in which the $SU(N_c)$ generators
carrying open indices $a_i$ appear in the order specified in the brackets. In the large-$N_c$ limit, interference terms between different
contributions to the amplitude vanish, and the notation  $\left[a\, ...\right]$ provides an efficient labeling of the contributions to the
gluon radiation cross section. 

The configuration $\left[a\right]$ arises for processes in which the final $q$-$g$ pair is in the same (total) color configuration as the incoming quark.
For a quark produced inside the medium, this includes the entire $0$-th order opacity contribution,  as well as a negative probability conserving term to order $N=1$. For a quark produced in the distant past, energy-momentum conservation ensures that the contribution $\left[a\right]$ is absent to orders $N=0$ and $N=1$ in opacity, but it is found to contribute to order $N=2$ with positive weight. 

The other vacuum-like color configurations $ \left[a\, a_1\right]$ , $\left[ a\, a_2\, a_1\right]$, and more generally configurations of the form
$\left[ a\, a_n\, a_{n-1}\, \dots a_1\right]$, arise first to $n$-th order in opacity as real terms with $n$ non-vanishing momentum and color
exchanges between medium and projectile parton. Higher orders in opacity will then provide probability-conserving corrections to the
same color configuration. Such corrections contain virtual terms, as exemplified by  the $N=2$ opacity contribution to $\left[a\, a_1\right]$. 

In general, to each order in opacity, only one vacuum-like color configuration $\left[ a\, a_n\, a_{n-1}\, \dots a_1\right]$ opens
up anew. However, with increasing order in opacity, more and more medium-modified color configurations can arise. For instance, up to second 
order in opacity, we find the terms 
\begin{equation}
	 \left[a_1\, a\right]\, ,\qquad \left[ a_2\, a\, a_1\right]\, , \qquad \left[a_2\, a_1\, a\right]\, ,\qquad  \left[ a_1\, a\, a_2\right]\, , \qquad \left[a_1\, a_2\, a\right]\, .
\end{equation}
The strong increase in the number of medium-modified color configurations with increasing order in opacity, and the corresponding decreasing weight of vacuum-like contributions (as seen in the explicit $N=2$ calculation) both support the physically intuitive idea that with increasing density of the medium it becomes easier for a radiated gluon to decohere in color from its partonic sister fragments. Our calculations show that the contribution of color-decohered gluons to the medium-induced radiation always exceeds 50 \% whenever the latter is sizable, and it may well be much larger than 50 \%.

\section{Hadronizing parton showers with medium-modified color flow}
\label{sec5}
Since QCD conserves color, a dynamically consistent hadronization model must respect color flow in interfacing the fragments of perturbative parton showers with the hadronic final states. Thus, even if hadronization occurs time-delayed and outside the medium, it can be affected by the modification of the color connections introduced by the medium.
The \textsc{Lund} string-fragmentation model is a prime example of a phenomenologically successful hadronization model that respects color flow. It is very well documented in the literature~\cite{Andersson:1998tv,Sjostrand:2006za}. Hadronization is modeled by decomposing each parton shower into a set of color singlet `strings'. These strings stretch between $q$ and $\bar{q}$ endpoints and represent color-connected gluons as kinks. They are then decayed into hadrons through the excitation of $q\bar{q}$ pairs from the vacuum.
In section~\ref{sec2} we already discussed some examples of interfacing the final stage of a branching process with \textsc{Lund} strings.
Since the medium-modified parton branchings studied in sections~\ref{sec3} and ~\ref{sec4} define color flow unambiguously, the above string-fragmentation routine can be applied without additional model assumptions. In this section, we explore how the distribution of hadronic fragments can be affected by medium-modifications of color flow.
To this end, we interface the possible color-configurations arising from the medium-induced branching of high-energy partons in the plasma with the \textsc{Lund} hadronization model. Numerical results were obtained with the string-fragmentation routine implemented in \textsc{Pythia 6.4}~\cite{Sjostrand:2006za}. The strings to be decayed were built using the routines PY1ENT (to add a given parton to the event) and PYJOIN (to join the partons provided by the user into a string, according to the proper order).   
The aim of the following analysis is to illustrate, for specific examples, how the distribution of hadronic fragments can be affected by a medium-modified color flow.

\begin{figure}[!htp]
\begin{center}
\includegraphics[clip,width=0.4\textwidth]{quark1_string.eps}
\hskip 1cm
\includegraphics[clip,width=0.48\textwidth]{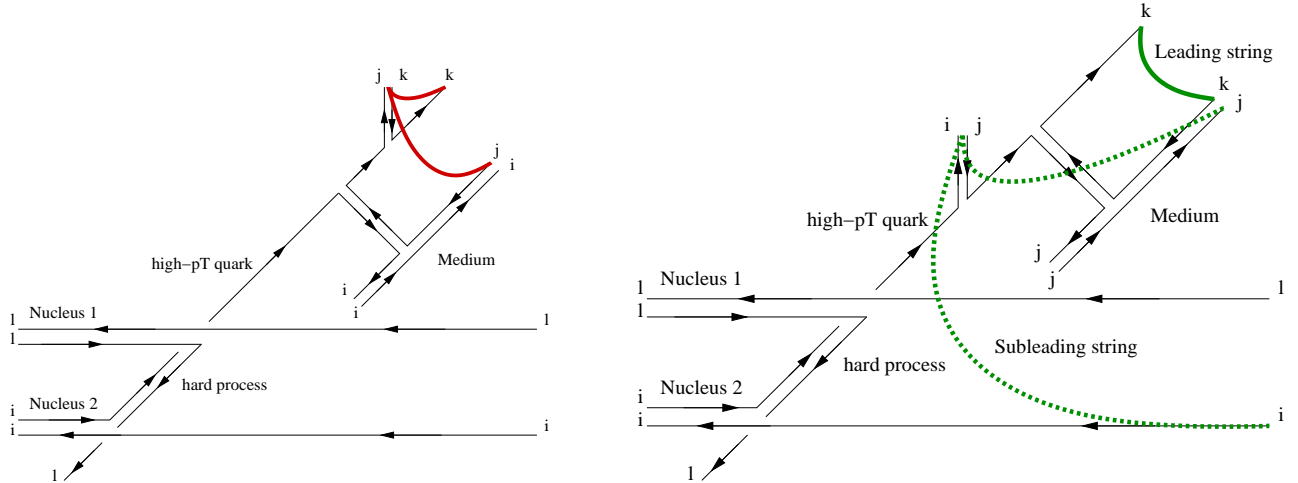}
\caption{The $N=1$ opacity correction to the fragmentation of a high-$p_T$ quark, interfaced with the \textsc{Lund}-string hadronization routine.
In the case of final state radiation (FSR, left hand side) the radiated gluon remains color connected with the other daughter of the branching and, if sufficiently collinear, can contribute to the energy of the leading hadron.
For initial state radiation (ISR, right hand side), the gluon results color decohered from the leading projectile fragment, independently on the emission angle; an independent Lund string is associated to it, whose decay will contribute to an enhanced multiplicity of soft particles.} 
\label{fig10}
\end{center}
\end{figure}

We start by considering Fig.~\ref{fig10}. To first order in opacity, medium-induced gluon radiation of a quark projectile exhibits two different color flows.
In the channel labeled as Final State Radiation, the emitted gluon will be part of the same \textsc{Lund} string as the leading quark.
In contrast, in the case of Initial State Radiation (right panel of Fig.~\ref{fig10}), the radiated gluon is color decohered from the projectile fragment; this means that it is instead linked 
through an independent string to a low-$p_T$ particle in the medium. 
For the following, the \textsc{Lund} strings in Fig.~\ref{fig10} are defined in terms of the 4-momenta of their end-points and kinks.
For simplicity both the leading quark fragment and the radiated gluon are assumed to be emitted at mid-rapidity ($\eta = 0$, i.e. $\theta = \pi/2$) and at relative azimuthal angle $\phi$. The other endpoint of the string will 
then be attached either to a particle from the medium or from the beam remnant: in both cases it will sit at low-$p_T$. Medium particles are taken with a typical thermal energy $\sim 3T$, with random event-by-event momentum orientation.
For the temperature we take the value $T=200$ MeV\footnote{The results were found to have a negligible  dependence on the temperature. This is consistent with the idea that the precise position of the soft endpoint of a string is unimportant for the hadronization of hard fragments.}. 
For the antiquark from the beam remnant, we choose a large momentum along the beam direction ($E=1$ TeV and $p_T=0$).
The only dependence on $E$ resides in the extension of the plateau in the rapidity distribution of (very soft) hadrons from the decay of the subleading string. 

\begin{figure}[!htp]
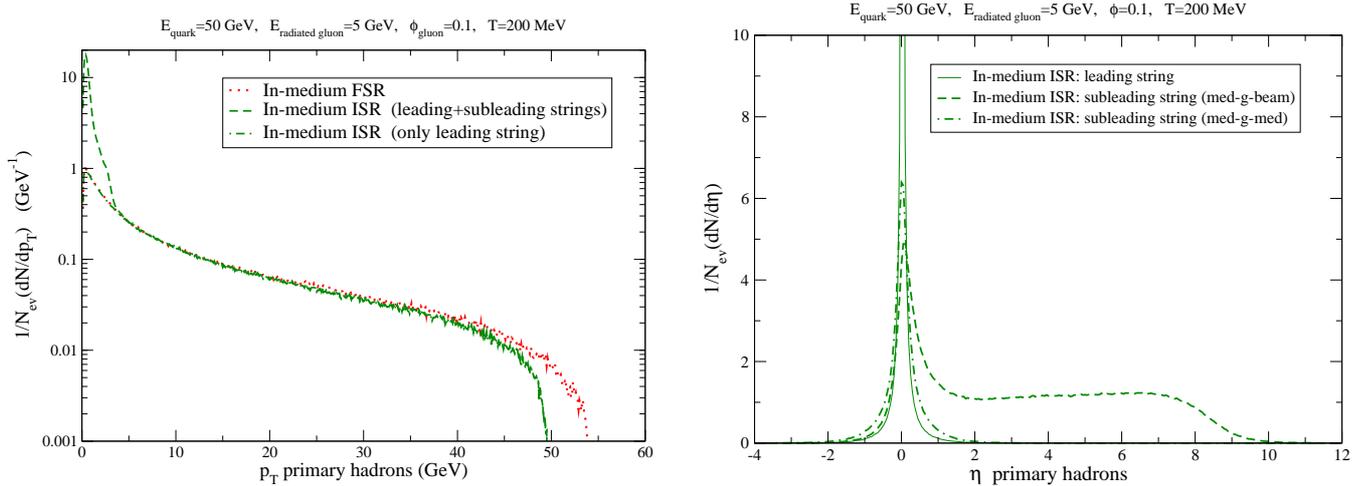

\begin{center}
\includegraphics[clip,width=0.48\textwidth]{FF_50+5_nobox.eps}
\hskip .5cm
\includegraphics[clip,width=0.48\textwidth]{dNdeta_50_ISR_medvsbeam.eps}
\caption{The $p_T$ and $\eta$ distributions  of the hadrons from the fragmentation of the Lund strings shown in Fig.~\protect\ref{fig10}. Both the quark and the gluon are emitted at midrapidity at relative angle $\phi = 0.1$. Left panel: fragmentation pattern in the FSR (in red) and ISR (in green) color channels. Right panel: rapidity distribution of the hadrons in the ISR channel. The sharpest peak around to $\eta=0$ (continuous line) comes from the fragmentation of the leading string. The pattern ``broad peak + plateau'' (dashed line) arises from the fragmentation of the subleading string, connected to the beam remnant (hence the long plateau). Also shown (dot-dashed line) is the case in which both endpoints of the subleading string are attached to a medium particle.} 
\label{fig11}
\end{center}
\end{figure}

Figure~\ref{fig11} shows examples of the distribution of hadronic fragments of the \textsc{Lund} strings depicted in Fig.~\ref{fig10}. These results were obtained
for a typical partonic configuration with a quark at high transverse momentum $p_T$ and a radiated gluon at much smaller transverse momentum (as an illustration we chose here and in the following $k_T = 0.1\, p_T$). 
In the left panel of Fig.~\ref{fig11} we display, for the two different color channels,  the fragmentation pattern of a hard quark branching in the medium.
The FSR case (red curve) corresponds to a vacuum-like color flow: in this case there is hadronic yield in a transverse momentum range that exceeds the $p_T$ of the leading quark. In the \textsc{Lund} model, this accounts for the fact that QCD is a finite resolution theory in which a perturbatively radiated gluon does not automatically increase the hadronic multiplicity by order unity or more: it is not necessarily `lost' but, remaining color-connected with the other daughter of the branching, may still contribute to the formation of the leading hadron.
In contrast, the ISR case (green curve) clearly shows that medium modification of color connections between the radiated gluon and the projectile fragment results in a softening of the hadron distribution: all hadronic yield above $p_T$ is suppressed and an additional contribution arises at soft momenta below $k_T$. The reason is that, for the ISR contribution, the color-decohered gluon and quark belong to different strings and thus cannot contribute to the same leading hadronic fragment. Therefore, hadronic multiplicity increases by construction with each color-decohered gluon by order unity or more, and the additional multiplicity is found in soft fragments  of transverse momentum lower than $k_T$, which is much smaller than $p_T$.
 
 These differences in the color flow of the ISR and FSR contribution have consequences for the distribution of hadronic fragments. In particular, the fragmentation of the \textsc{Lund}
 string of a vacuum-like (FSR) contribution results mainly in semi-hard and hard hadrons. For instance, fragmentation of the FSR string of total energy $\sim 55$ GeV in Fig.~\ref{fig11}  
 yields on average $\langle N_h\rangle=5.4$ hadrons, of which 3.9 carry $p_T > 2$ GeV transverse momentum. Since the multiplicity of \textsc{Lund} strings grows only mildly with
the total length and with the number of small kinks, the string of the ISR contribution that contains the leading quark fragment will decay into almost as many hadrons ($\langle N_h\rangle=5.2$
 for the case shown in in Fig.~\ref{fig11}). However, these hadrons carry a smaller total transverse momentum, the remaining fraction $k_T/ \left( k_T + p_T\right)$ being
 carried by the fragments of the subleading string. The latter stretches over a short distance in transverse momentum, but it can stretch over a long distance in rapidity, and 
 hence it can yield high multiplicity. For instance, for one of the cases shown in Fig.~\ref{fig11}, the string stretches from a medium component around $\eta =0$ to a beam
 remnant at projectile rapidity and thus distributes $\langle N_h\rangle_{\rm sublead}^{\rm med-g-beam}=12.7$ hadrons over approximately 10 units in rapidity. Alternatively,
it is conceivable that due to further interactions between the medium and the radiated gluon, both ends of the subleading string connect to soft components close to mid-rapidity. 
In this case, the subleading string produces a much smaller number of hadrons ($\langle N_h\rangle_{\rm sublead}^{\rm med-g-med}=3.7$ for the case in Fig.~\ref{fig11});
the hadronic distribution close to the rapidity $\eta = 0$ of the jet,  however, changes only mildly.

Recent measurements by the CMS collaboration indicate that the medium-modified fragmentation of jets in heavy ion collisions at the LHC is accompanied by a significant increase in soft hadronic multiplicity~\footnote{An increase in soft hadronic multiplicity is also seen in hadron- and photon-triggered jet-like distributions at RHIC. For recent preliminary data, see e.g. Refs.~\cite{Ohlson:2011qr,Connors:2011zz}.} that appears outside typical jet cones at large angle with respect to the jet axis~\cite{Chatrchyan:2011sx}, while the hard part of the jet fragmentation function is consistent with that of a vacuum jet of lower transverse energy \cite{cms:ff}. Here, we note that medium-modified color decoherence is a natural candidate mechanism for two of these qualitative features. Namely, the leading string of 
the ISR contribution coincides by construction with that of a vacuum quark jet of lower transverse energy ($p_T$, rather than $p_T + k_T$). Moreover,  medium-induced color 
decoherence leads naturally to distributing the transverse energy $k_T$ amongst very soft components. It is less clear, however, whether this additional soft multiplicity is distributed 
naturally over large angles outside the jet cone. Our study indicates that the color decoherence associated to the $N=1$ opacity calculation is not enough to achieve such an angular broadening and a larger amount of rescattering at the partonic level (that can be only achieved going to higher orders in opacity) is mandatory. We caution that the numerical
studies presented here focused entirely on the effects of color decoherence while a phenomenologically satisfactory description of jet quenching requires of course
to interface these effects with the medium-modification of kinematic distributions. Consistent with previous discussions we therefore expect that the broadening of soft
components is mainly due to kinematic effects (see Refs.~\cite{CasalderreySolana:2010eh,Qin:2010mn,Young:2011qx,Linders:2011qm}  for models consistent with this idea). 

\begin{figure}[!htp]
\begin{center}
\includegraphics[clip,width=0.48\textwidth]{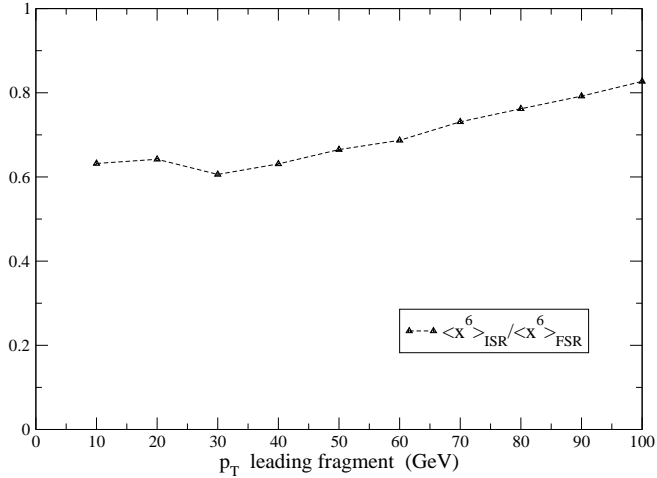}
\caption{The ratio of $6$-th moments of the ISR and FSR contributions to the fragmentation functions of partonic configurations shown in Fig.~\ref{fig10}. Here, prior to hadronization the leading quark has transverse momentum $p_T$, and the radiated gluon carries a scaled transverse momentum $\kg = 0.1\, p_T$ and is emitted at $\phi = 0.1$.} 
\label{fig12}
\end{center}
\end{figure}

We finally comment  on the potential relevance of our findings for single-hadron spectra which are particularly sensitive to the modifications occurring to the hard tail of the fragmentation function. It is customary to characterize this effect by considering a steeply-falling parton spectrum $1/p_T^n$ that if convoluted with a Fragmentation Function (FF) yields
 $d\sigma^{\rm hadron}\sim\langle x^{n-1}\rangle/p_T^n$. This shows that single hadron spectra are mainly sensitive to a higher ($(n-1)^{\rm th}$) moment of the FF.
In Fig.~\ref{fig11},  we display the ratio of the 6-th moments for different color channels in a specific kinematic configuration of a parton shower. This figure illustrates that 
for an identical kinematic partonic configuration (i.e., for the same amount of energy loss at the partonic level), the modification of the color connections can introduce
an additional significant source of suppression of the hadronic spectrum. Quantitatively, in the case displayed in Fig.~\ref{fig11}, one finds for instance $\langle x^6\rangle_{\rm ISR}=0.052$ and $\langle x^6\rangle_{\rm FSR}=0.078$ for the ISR and FSR channels respectively.
In summary: since medium-induced color decoherence of radiated gluons depletes naturally the high-$p_T$ tail of fragmentation functions, it can contribute to 
reducing the nuclear modification factor $R_{\rm AA}$. A similar observation was made in  a recent study~\cite{Beraudo:2011bh}, in which parton splittings with medium-modified color flow were interfaced with a cluster hadronization model. Also in that case the suppression showed a weak dependence on transverse momenta and persisted at transverse momenta exceeding 100 GeV.
Fig.~\ref{fig12} supports the picture that medium-induced color decoherence can be a relevant factor for understanding the value and $p_T$-dependence of the nuclear modification factor at high transverse momentum.

\begin{figure}[!htp]
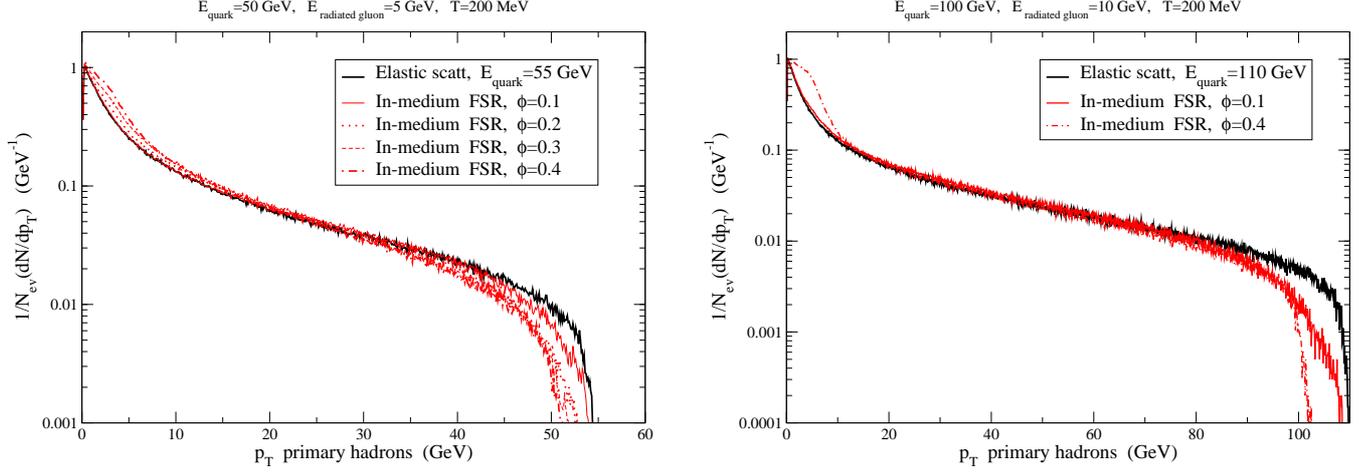

\begin{center}
\includegraphics[clip,width=0.48\textwidth]{vselastic_50.eps}
\hskip .5cm
\includegraphics[clip,width=0.48\textwidth]{vselastic_100.eps}
\caption{The same as Fig.~\ref{fig11}, but for two choices of quark energy, and for a varying azimuthal angle $\phi$ of the radiated gluon. As the relative angle gets larger the FF suffers a softening.} 
\label{fig13}
\end{center}
\end{figure}

\begin{figure}[!h]
\begin{center}
 \includegraphics[clip,width=0.48\textwidth]{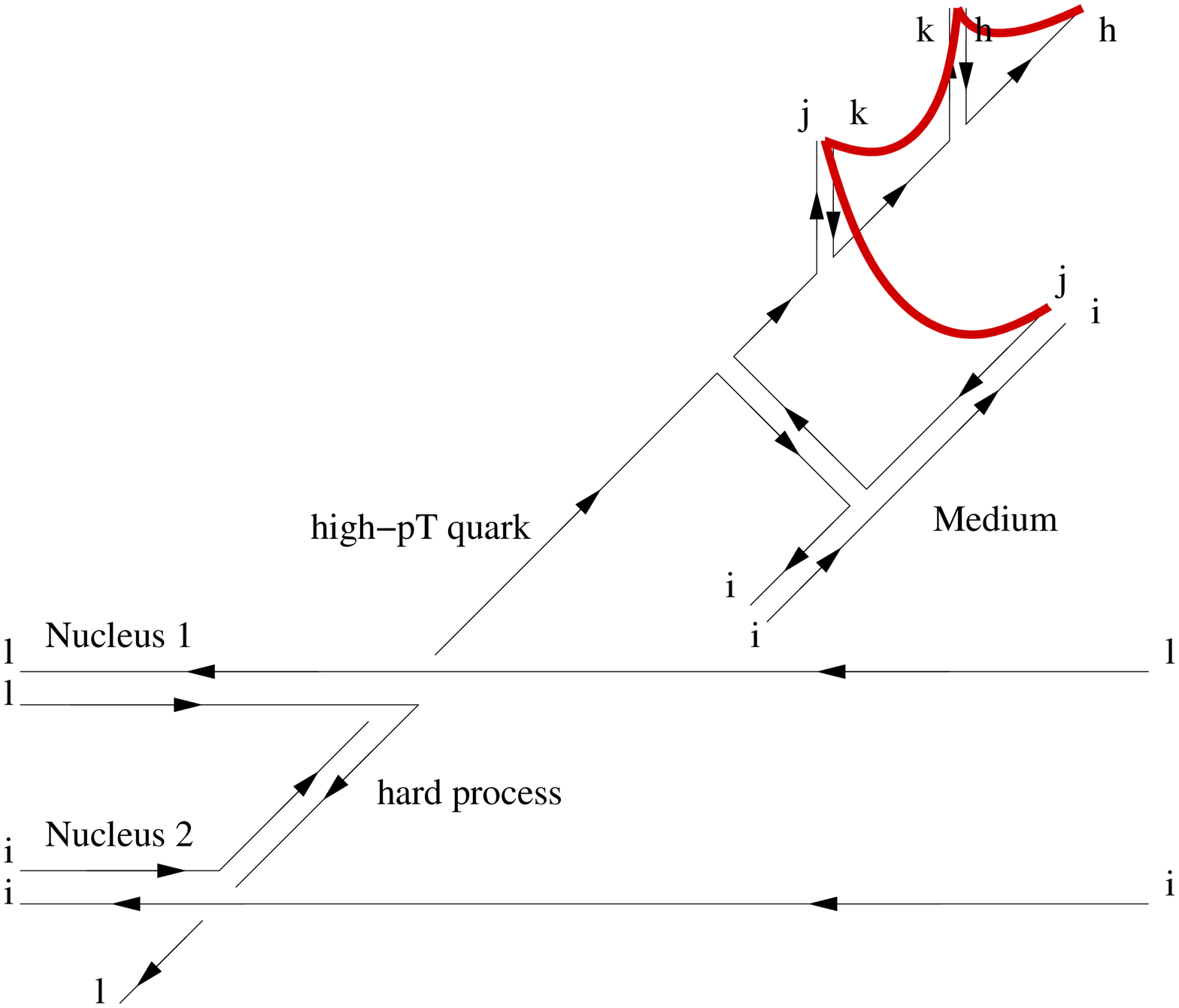}
 \hskip 1cm
\includegraphics[clip,width=0.4\textwidth]{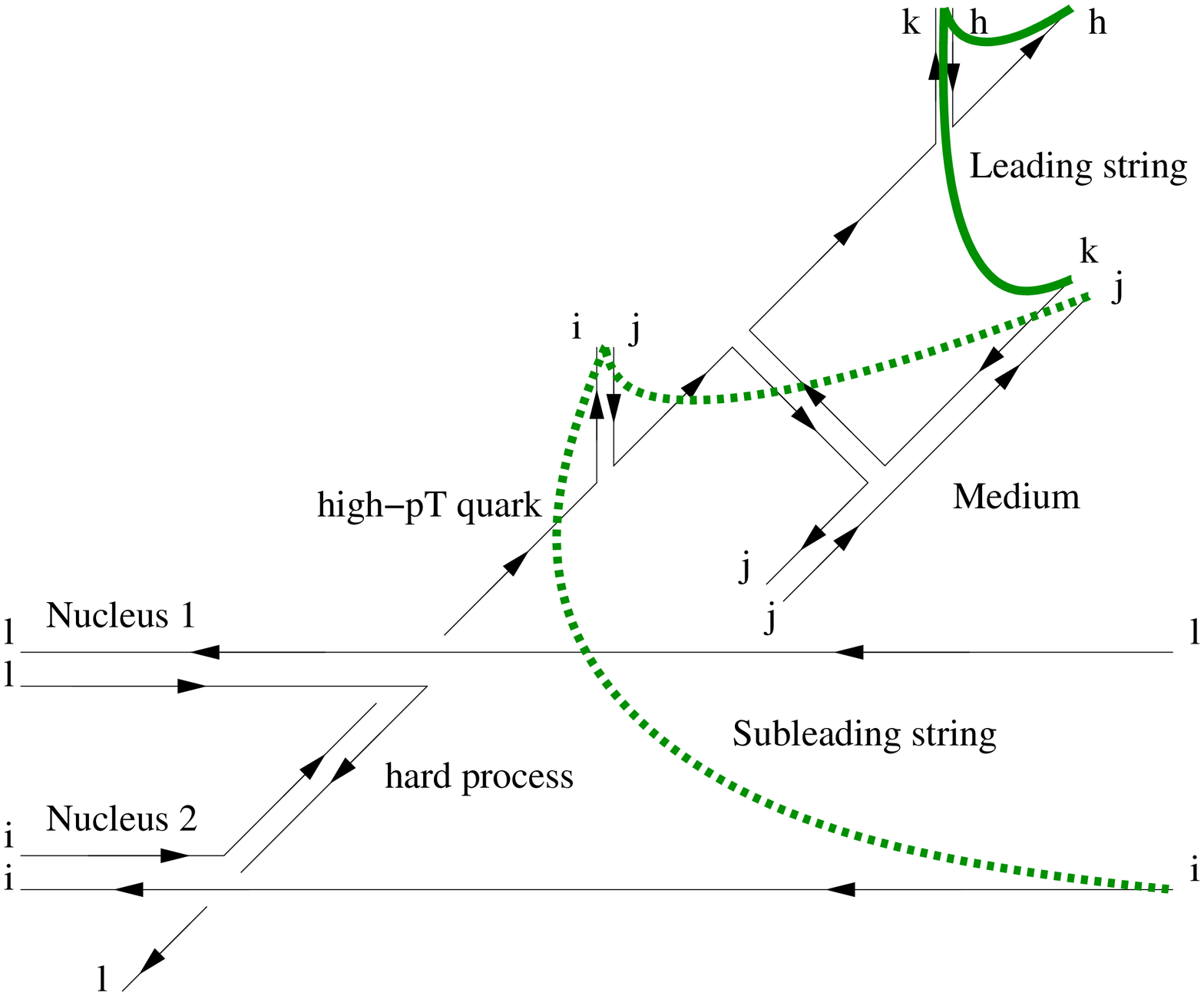}
\caption{The color flow associated with medium-induced FSR (left hand side) and ISR (right hand side) in the $N=1$ opacity expansion, supplemented with an additional second gluon emission at late time.}
\label{fig14}
\end{center}
\end{figure}

\begin{figure}[!h]
\begin{center}
\includegraphics[clip,width=0.48\textwidth]{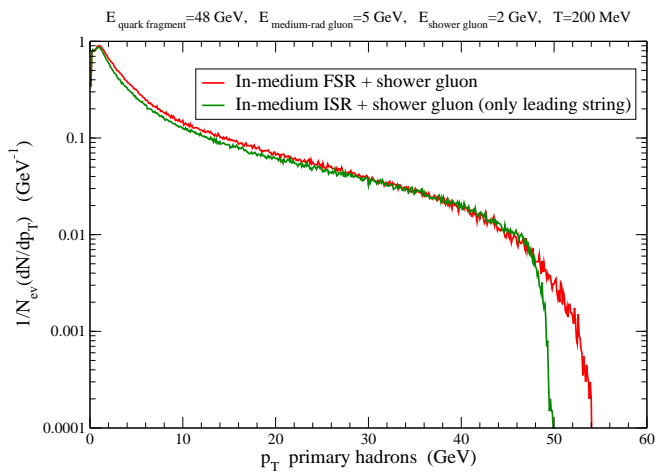}
\caption{The $p_T$-distribution of hadrons from the fragmentation of the Lund strings interfaced with the two color-flows involved at $N=1$ order in opacity supplemented by the subsequent showering shown in Fig.~\ref{fig14}. The two parton branchings were at azimuthal angle $\phi = 0.2$ and $\phi = 0.1$, respectively. The figure illustrates that the effects of medium-induced color decoherence are not washed out by subsequent showering. }
\label{fig15}
\end{center}
\end{figure}

Elementary formation time arguments suggest~\cite{CasalderreySolana:2011gx}
that hadronization can occur time-delayed outside the medium, and that the earliest parton branchings in a shower can be medium-modified since they occur inside the medium. However, these earliest branchings may be followed by subsequent partonic branchings outside the medium prior to hadronization. The question arises whether the features observed in Fig.~\ref{fig11} and ~\ref{fig12} are robust against inclusion of such subsequent parton branchings. 
The structure of typical parton branching histories indicates that inclusion of a single additional splitting on top of the ISR and FSR contributions of Fig.~\ref{fig10} provides already significant insight into the question of how final state branchings may alter the effects of medium-induced color decoherence on hadronization~\footnote{
As a consequence of the steeply falling $p_T$ dependence of spectra, by triggering on the $p_T$ of a parton at the end of a shower evolution one typically selects parton showers 
that have suffered a very limited number of branching. For instance, for a sample of quarks at LHC energy generated by \textsc{Pythia} in a `hard event' + `final state shower' process, 
one finds for a trigger $p_T\sim 50$ GeV of the final quark an average number of daughter gluons of order $\sim 2.5$. This number increases only very weakly with $p_T^{\rm trig}$.}. 
As shown in Fig.~\ref{fig14}, inclusion of another parton splitting adds an additional kink to the leading string, but it does not change the fact that the color-decohered gluon of the ISR configuration hadronizes in a separate, subleading string, while in the FSR contribution both gluons are part of the leading string. As a consequence, if both gluons of the FSR contribution are produced sufficiently collinear, they can contribute to the production of the leading hadron, and the corresponding fragmentation function in Fig.~\ref{fig15} extends up to the transverse momenta that correspond to the sum of the transverse momentum of the three projectile partons in the final state. In contrast, the color-decohered gluon of the ISR contribution is lost for the formation of the leading hadron. In this way, Figure~\ref{fig15} clearly illustrates that additional parton branching at late times does not wash out the effects of medium-induced color decoherence. More precisely, medium-induced color decoherence removes efficiently a fraction of a jet's energy from the high-$p_T$ part of the fragmentation function by hadronizing it independently in soft fragments. The leading string, however, irrespective of the number of gluons (i.e. kinks) radiated outside the medium that it involves, appears to correspond at sufficiently high $p_T$ to a vacuum fragmentation pattern of a jet of lower (ISR gluons are decohered) total transverse energy.
We emphasize that this observation depends mainly on two generic features, namely that color-decohered gluons correspond to different strings and that additional branchings at late times correspond to (typically small) kinks on the same string and thus contribute to the distribution of leading fragments. 

\section{Conclusions}

In the absence of medium effects,  the only color correlation between a final state parton shower and the rest of a hadronic collision is the one between the color 
of the primary `parent' parton and some color-compensating beam remnant. 
Jet-medium interactions change this feature generically and characteristically. 
To date, very few studies \cite{Sapeta:2007ad,Leonidov:2010xf,Leonidov:2010he,CasalderreySolana:2011rz,Aurenche:2011rd,MehtarTani:2010ma,MehtarTani:2011tz,MehtarTani:2011gf} have addressed 
aspects of the role played by medium modifications of color flow. Most studies focus solely on medium-induced changes of the multiplicity and kinematic distribution of partons in the shower.
In the present work, 
we have pointed out that jet-medium interactions give rise
to qualitatively novel color correlations in which components of the parton shower are color-decohered from their sister partons in branching processes.

Let us summarize our findings in a formulation that is slightly different from the one adopted so far.
Physically, hadronization of parton showers amounts to a procedure that specifies the overlap between the color singlet components in the shower
and the physical Hilbert space of all hadronic wave functions corresponding
to experimentally accessible states. As is generally known and as we have illustrated for the \textsc{Lund} hadronization model in Fig.~\ref{fig13}, this mapping can result in
hadronic fragments that are more energetic than the most energetic partonic component that is hadronized. This is 
so, since a radiated partonic component -- as long as not decohered completely from its sisters in the branching process -- can display an overlap with the same
hadronic one-particle state as its sisters. In this sense, the energy radiated perturbatively in gluons is not automatically an energy lost for the formation of a hadron;
it is only `lost' if the gluon is decohered. In the vacuum, sister partons typically remain color connected and their decoherence therefore arises only kinematically, 
as a consequence of their relative transverse separation. To date, studies of parton energy loss have mainly focused on modifications of this kinematic decoherence.
Here, we have demonstrated that in comparison to purely
kinematic effects, medium-modified color flow is a highly efficient decoherence mechanism and can thus contribute significantly to softening 
hadronic distributions (see section~\ref{sec5}). As emphasized in section~\ref{sec2} this is a qualitatively novel and generic property 
of parton energy loss mechanisms that arises  from the gluon exchanges between the parton shower and the medium, independently of model-specific dynamical 
details.  It will generically increase the number of decohered color singlets in the parton shower. The relative rate of partonic configurations with an increased number of color-decohered components
is large: in the explicit color-differential calculations presented in sections~\ref{sec3} and ~\ref{sec4}, it always exceeds 50 \% of the entire medium-modified 
gluon radiation.

Remarkably, we found in section~\ref{sec5} that the medium-induced color decoherence of gluons from the parton shower leads naturally to their hadronization 
into {\it soft} components while the most energetic components will hadronize like vacuum structures of reduced transverse energy. This appears to be characteristically 
different from the typically studied kinematically induced medium modifications where one expects generically that the enhancement 
of soft jet fragments is accompanied by medium-modifications of the shape of the fragmentation function at all momentum scales. On the other hand, it is
qualitatively in line with CMS data on dijet asymmetries. In particular, CMS data show jet fragmentation functions that are at high transverse momentum consistent with 
the fragmentation of vacuum jets of degraded total transverse energy $E_T - \Delta E$. Moreover,  the missing jet energy $\Delta E$ is recovered in very soft 
hadrons. Both these features arise naturally from the medium-modified color decoherence of jet fragments studied here. We caution, however, that the experimentally 
observed soft components are distributed over a wide region in $\Delta\eta \times \Delta \phi$. While we have 
seen some broadening of the distribution of soft fragments (see Fig.~\ref{fig13}), we did not identify a generic argument that could account for such an observed
wide distributions in the presence of only $N=1$ gluon exchange with the medium. It remains to be clarified for instance to what extent soft fragments are broadened 
further if subsequent  medium-induced gluon branchings or higher orders in opacity are taken into account. More generally, the present work was limited to
showing that medium-induced color decoherence is a non-negotiable aspect of parton energy loss mechanisms that can lead to  
qualitatively novel and numerically significant features in jet quenching calculations. The next step is now to incorporate
the generic features of medium-induced color decoherence established here in a more complete dynamical modeling of jet quenching.

\section*{Acknowledgements}

We thank N.  Armesto, C. Salgado, B.G. Zakharov and K. Zapp for useful discussions.
JGM acknowledges the support of Funda\c c\~ao para a Ci\^encia e a Tecnologia (Portugal) under project CERN/FP/116379/2010. 
\appendix

\section{Gluon projectile: color-differential spectrum}
\label{appa}
Here we provide further details on the color-differential $N=1$ opacity calculation for the case of a projectile gluon impacting from the distant past. 
This extends the discussion of  the diagrams of Fig.~\ref{fig6} in section~\ref{sec3A} to the case of a medium-modified splitting
\beq
 g(d) \to g(b)\,  g(a)\, .
\eeq
As indicated above, the incoming gluon has color $d$. The outgoing gluons carry momentum fractions $\xg$ and $1-\xg$ with $\xg\ll 1-\xg$. The color $b$
is attributed to the outgoing harder gluon that carries momentum fraction $1-\xg$ and to which we refer as outgoing `projectile' gluon; the softer outgoing gluon carries color $a$ and the gluon exchanged with the medium has color index $a_1$. 
The polarization vectors for the incoming and outgoing `projectile' gluon, as well as for the radiated gluon are
\beq
\ei\equiv[0,0,\eei],\quad\ef\equiv\left[0,\frac{\eef\!\cdot\!(\q\!-\!\kkg)}{(1\!-\!\xg)p_+},\eef\right] \, ,\quad
 \eg\equiv\left[0,\frac{\eeg\!\cdot\!\kkg}{\xg p_+},\eeg\right]\, ,
\eeq
respectively. For a high energy gluon, the interaction with the medium is described by the eikonal vertex $g f^{abc}g^{\mu\rho}(p+p')^\nu$.
The amplitude ${\cal M}_{(a)}$ of a gluon that first interacts with the medium prior to gluon radiation reads then
\beqa
i{\cal M}_{(a)} &=& \sum_{n=1}^{N} g f^{abc}g^{\nu\rho}(-2p_f-\kg)^\mu\frac{(-i)}{(p_f+\kg)^2}\times 
g f^{ca_1d} g^\nu_\rho(2p_i+q)^\sigma\epsilon_{g,\mu}\epsilon_{f,\nu}\epsilon_{i,\eta}A_\sigma(q)e^{iq\cdot x_n}T^{a_1}_{(n)}
	\nonumber \\
&=& \sum_{n=1}^{N}(-i)\,g^2\left(T_A^aT_A^{a_1}\right)_{bd}(\eei\!\cdot\!\eef)\left(\frac{p_f\!\cdot\!\eg}{p_f\!\cdot\!\kg}\right)\,2p^+{\cal A}(\q)e^{iq\cdot x_n}T^{a_1}_{(n)}.
\eeqa
In close analogy, we find
\beq
i{\cal M}_{(b)}=\sum_{n=1}^{N}-(-i)\,g^2\left(T_A^{a_1}T_A^{a}\right)_{bd}(\eei\!\cdot\!\eef)\,(1\!-\!\xg)\left(\frac{p_i\!\cdot\!\eg}{p_i\!\cdot\!\kg}\right)\,2p^+{\cal A}(\q)e^{iq\cdot x_n}T^{a_1}_{(n)}\, ,
\eeq
and
\beq
i{\cal M}_{(c)}=\sum_{n=1}^{N}(-i)g^2 f^{aa_1c}f^{bcd}(\ei\!\cdot\!\ef)\frac{-1}{(\kkg\!-\!\q)^2}
2(1\!-\!\xg)\,[\eeg\!\cdot\!(\kkg\!-\!\q)] 2p^+ {\cal A}(\q)e^{iq\cdot x_n}T^{a_1}_{(n)} \, .
\eeq
These expressions are given in terms of the generators of the adjoint representation, $\left(T^a_A\right)_{bc} = i f^{abc}$. For a discussion of color flow in the large-$N_c$ limit it is useful to express them in terms of traces over products of generators $t^a$ in the fundamental representation. For instance, starting from the normalization ${\rm Tr}(t^at^b)\!=\!1/2\delta^{ab}$ and $i f^{abc} = 2\, {\rm Tr} \left(\left[t^a,t^b\right] t^c \right)$, one finds
\beq
(T^aT^{a_1})_{bd}=2\,{\rm Tr}\left([t^b,t^a][t^{a_1},t^d]\right).
\eeq
As explained in section~\ref{sec3b}, this allows one to separate the entire gluon radiation amplitude into six different color channels. The corresponding diagrammatic contributions are shown in Fig~\ref{fig5} and listed in (\ref{eq3.15}). Here, we provide their explicit expressions. Representing the incoming gluon projectile in the large-$N_c$ limit by a quark and an anti-quark leg, we label these contributions as follows:
\begin{itemize}
\item FSR($q$): final state radiation of the quark leg (configuration $baa_1d$)
\beq
i{\cal M}_{baa_1d}=\sum_{n=1}^{N}(-i)\,2{\rm Tr}\left(t^bt^at^{a_1}t^d\right)\,\left(\frac{\eeg\!\cdot\!(\kkg\!-\!\xg\q)}{(\kkg\!-\!\xg\q)^2}-\frac{\eeg\!\cdot\!(\kkg\!-\!\q)}{(\kkg\!-\!\q)^2}\right)
 2(1\!-\!\xg) (\eei\!\cdot\!\eef){\cal M}_{\rm el}^{a_1,(n)}
\eeq
\item FSR($\bar{q}$): final state radiation of the anti-quark leg (configuration $bda_1a$)
\beq
i{\cal M}_{bda_1a}=\sum_{n=1}^{N}(-i)\,2{\rm Tr}\left(t^at^bt^{d}t^{a_1}\right)\,\left(\frac{\eeg\!\cdot\!(\kkg\!-\!\xg\q)}{(\kkg\!-\!\xg\q)^2}-\frac{\eeg\!\cdot\!(\kkg\!-\!\q)}{(\kkg\!-\!\q)^2}\right)
 2(1\!-\!\xg) (\eei\!\cdot\!\eef){\cal M}_{\rm el}^{a_1,(n)}
\eeq
\item ISR($q$): initial state radiation of the quark leg (configuration $ba_1ad$)
\beq
i{\cal M}_{ba_1ad}=\sum_{n=1}^{N}-(-i)\,2{\rm Tr}\left(t^bt^{a_1}t^{a}t^{d}\right)\,\left(\frac{\eeg\!\cdot\!\kkg}{\kkg^2}-\frac{\eeg\!\cdot\!(\kkg\!-\!\q)}{(\kkg\!-\!\q)^2}\right)
 2(1\!-\!\xg) (\eei\!\cdot\!\eef){\cal M}_{\rm el}^{a_1,(n)}
\eeq
\item ISR($\bar{q}$): initial state radiation of the anti-quark leg (configuration $bdaa_1$)
\beq
i{\cal M}_{bdaa_1}=\sum_{n=1}^{N}-(-i)\,2{\rm Tr}\left(t^{a_1}t^{b}t^{d}t^{a}\right)\,\left(\frac{\eeg\!\cdot\!\kkg}{\kkg^2}-\frac{\eeg\!\cdot\!(\kkg\!-\!\q)}{(\kkg\!-\!\q)^2}\right)
 2(1\!-\!\xg) (\eei\!\cdot\!\eef){\cal M}_{\rm el}^{a_1,(n)}
\eeq
\item RqS$\bar{q}$: radiation of the quark leg but medium interaction on anti-quark leg, configuration ($bada_1$)
\beq
i{\cal M}_{bada_1}=\sum_{n=1}^{N}-(-i)\,2{\rm Tr}\left(t^bt^at^{d}t^{a_1}\right)\,\left(\frac{\eeg\!\cdot\!(\kkg\!-\!\xg\q)}{(\kkg\!-\!\xg\q)^2}-\frac{\eeg\!\cdot\!\kkg}{\kkg^2}\right)
 2(1\!-\!\xg) (\eei\!\cdot\!\eef){\cal M}_{\rm el}^{a_1,(n)}
\eeq
\item R$\bar{q}$Sq: radiation of the anti-quark leg but medium interaction on quark leg (configuration $ba_1da$)
\beq
i{\cal M}_{ba_1da}=\sum_{n=1}^{N}-(-i)\,2{\rm Tr}\left(t^at^bt^{a_1}t^d\right)\,\left(\frac{\eeg\!\cdot\!(\kkg\!-\!\xg\q)}{(\kkg\!-\!\xg\q)^2}-\frac{\eeg\!\cdot\!\kkg}{\kkg^2}\right)
 2(1\!-\!\xg) (\eei\!\cdot\!\eef){\cal M}_{\rm el}^{a_1,(n)}
\eeq
\end{itemize}
In the above, the first four channels are consistent with the simplified large-$N_c$ picture, where  the radiation from a gluon is viewed as the incoherent superposition of the medium-induced radiation from its quark and anti-quark legs. The last two channels, on the other hand, represent contributions for which the $q$ ($\bar{q}$) leg radiates and the $\bar{q}$ ($q$) leg exchanges color with the medium. For these contributions, a distinction between initial state and final state radiation is not needed, since it does not affect the color ordering of the amplitude. However, in the soft $\xg\ll 1$ limit, these last two channels give a subleading contribution. Squaring these two amplitudes one would find a QED-like (`Bethe-Heitler') radiation spectrum that is $O(\xg^2)$. On the other hand, the first four channels are not suppressed in the soft limit and provide, to leading order in $\xg$, an identical contribution: they sum up to the well-known Gunion-Bertsch spectrum. 

\section{Details on the $N=2$ calculation}
\label{app:n2}
In section~\ref{sec4} we quoted the result for the color-differential radiation spectrum at order $N=2$ in opacity. The physical meaning of the various contributions became particularly transparent by expressing the result in terms of the effective currents $\J_i$ ($i=1,2,12$). Here we give details on how to derive these results. For this purpose, the calculation of a few diagrams will be sufficient, the other ones will follow from the modified dependence on the currents $\J_i$.   

\subsection{Direct contribution}
\begin{figure}[!htp]
\begin{center}
\includegraphics[clip,width=0.75\textwidth]{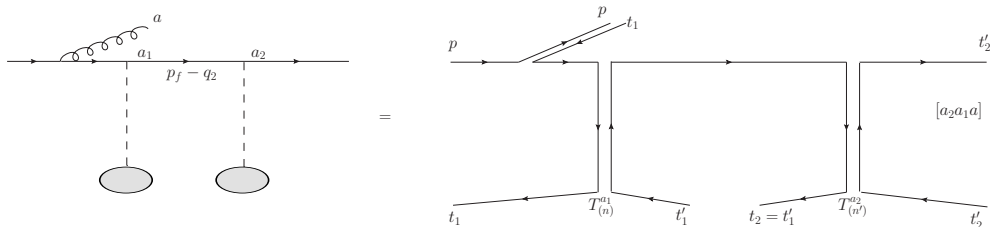}
\caption{A contribution to ${\cal M}_2^{\rm dir}$ with its corresponding `$a_2a_1a$' color flow (which gets contribution also from the diagram with the radiated gluon scattering on the first scattering center).}
\label{fig:N2initial}
\end{center}
\end{figure}
\begin{figure}[!htp]
\begin{center}
\includegraphics[clip,width=0.6\textwidth]{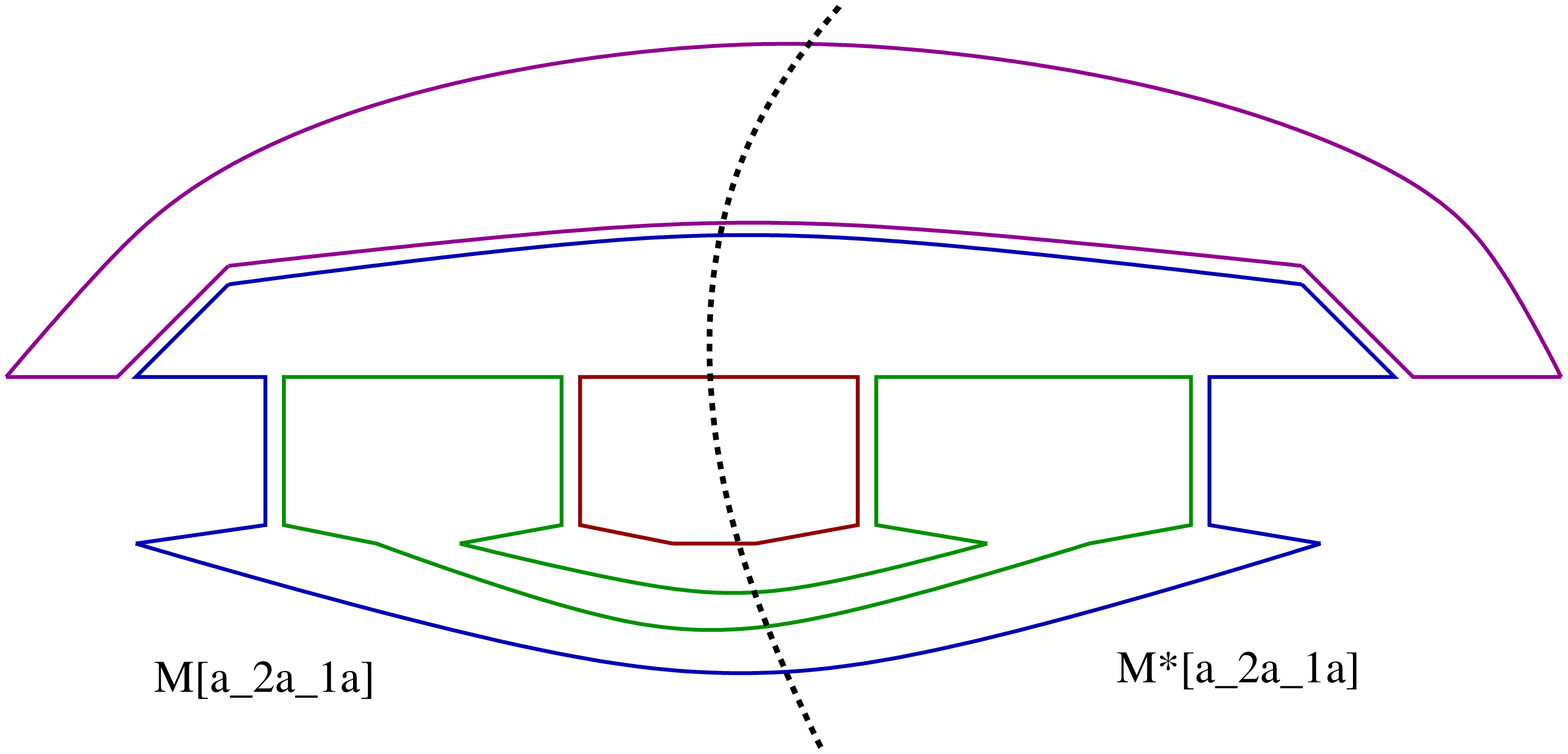}
\caption{The color flow in $\langle|{\cal M}_2^{a_2a_1a}|^2\rangle$. Each disconnected loop gives rise to a factor $N_c$. Each gluon line provides a factor $1/2$ (due to the Fierz identity).}
\label{fig:N2flow}
\end{center}
\end{figure}
We start by considering the direct term in the $N=2$ spectrum.  For illustration, we focus on the 
`$a_2a_1a$' channel that can be written as a sum of two terms, 
\beq
{\cal M}_2^{a_2a_1a}\equiv{\cal M}_{a_2a_1a}^{(I)}+i{\cal M}_{a_2a_1a}^{(II)}\, ,
\eeq
where ${\cal M}_{a_2a_1a}^{(I)}$ is the contribution with the gluon radiated before any interaction with the medium (see Fig.~\ref{fig:N2initial}) and ${\cal M}_{a_2a_1a}^{(II)}$ describes the radiated gluon interacting with the first scattering center. The first term takes the explicit form
\begin{multline}
i{\cal M}_{a_2a_1a}^{(I)}=\sum_{n_1,n_2}g(t^{a_2}t^{a_1}t^{a})(T_{(n_2)}^{a_2}T_{(n_1)}^{a_1})\int\frac{dq_2^-}{2\pi}\int\frac{d\q_1}{(2\pi)^2}\int\frac{d\q_2}{(2\pi)^2}e^{-i\q_1\cdot\x_{n_1}}e^{-i\q_2\cdot\x_{n_2}}\\
\times e^{i(q_1^-+q_2^-)x_{n_1}^+}e^{iq_2^-(x_{n_2}^+-x_{n_1}^+)}(i)[(2p_f-q_2)^+{\cal A}(\q_2)]\frac{i}{(p_f-q_2)^2-M^2+i\eta}\\
\times(i)[(2p_f-2q_2-q_1)^+{\cal A}(\q_1)]\frac{i}{(p_i-\kg)^2-M^2+i\eta}(i)[(2p_i-\kg)\cdot\eg].
\end{multline}
Here, the pole of the propagator ($\Q\!\equiv\!\q_1\!+\!\q_2$)
\beq
\frac{i}{(p_f-q_2)^2-M^2+i\eta}=\frac{-i}{2(1-\xg)p^+[q_2^--(\frac{(\Q-\kkg)^2-(\q_1-\kkg)^2}{2(1-\xg)p^+}+i\eta)]}
\eeq
allows one to perform the $q_2^-$ integration. Furthermore, momentum conservation implies
\beq
q_1^-+q_2^-=\frac{\kkg^2}{2\xg p^+}+\frac{(\Q-\kkg)^2+M^2}{2(1-\xg)p^+}-\frac{M^2}{2p^+} \underset{\xg\ll 1}{\sim}\wob\, .
\eeq
Up to ${\cal O}(\xg)$ corrections, one gets
\begin{multline}
i{\cal M}_{a_2a_1a}^{(I)}=-g\sum_{n_1,n_2}(t^{a_2}t^{a_1}t^{a})(T_{(n_2)}^{a_2}T_{(n_1)}^{a_1})\int\frac{d\q_1}{(2\pi)^2}\int\frac{d\q_2}{(2\pi)^2}e^{-i\q_1\cdot\x_{n_1}}e^{-i\q_2\cdot\x_{n_2}}\\
\times e^{i\wob x_{n_1}}\,{\cal A}(\q_1)\,{\cal A}(\q_2)\left(2\eeg\cdot\frac{\kkg}{\kkg^2+\xg^2M^2}\right)\,2p^+.
\end{multline}
Proceeding analogously with the contribution ${\cal M}_{a_2a_1a}^{(II)}$, one obtains
\begin{multline}
i{\cal M}_2^{a_2a_1a}=-g\sum_{n_1,n_2}(t^{a_2}t^{a_1}t^{a})(T_{(n_2)}^{a_2}T_{(n_1)}^{a_1})\int\frac{d\q_1}{(2\pi)^2}\int\frac{d\q_2}{(2\pi)^2}e^{-i\q_1\cdot\x_{n_1}}e^{-i\q_2\cdot\x_{n_2}}\\
\times e^{i\wob x_{n_1}}{\cal A}(\q_1){\cal A}(\q_2)\left(2\eeg\cdot\J_1\right)\,2p^+.
\end{multline}
After squaring, tracing over colors
\beq
\frac{1}{d_n}{\rm Tr}\left(T_{(n)}^{a_i}T_{(n')}^{a_{i'}}\right)=\frac{1}{d_n}\delta_{n,n'}T_n\delta^{a_ia_{i'}}\label{eq:colortrace}\, ,
\eeq
and averaging the phase factor over the transverse location of the scattering centers
\beq
\frac{1}{A_\perp}\int d\x_n e^{-i(\q_n-\q_{n'})\cdot\x_2}=\frac{1}{A_\perp}(2\pi)^2\delta(\q_n-\q_{n'})\, ,
\eeq
one finds
\beq
\langle|{\cal M}_2^{a_2a_1a}|^2\rangle=g^2C_F\sum_{n_1,n_2}\frac{1}{A_\perp}\frac{C_FT_{n_1}}{d_{n_1}}\!\!\!\int\!\!\frac{d\q_1}{(2\pi)^2}|{\cal A}(\q_1)|^2\frac{1}{A_\perp}\frac{C_FT_{n_2}}{d_{n_2}}\!\!\!\int\!\!\frac{d\q_2}{(2\pi)^2}|{\cal A}(\q_2)|^2\,4|\J_1|^2\,(2p_+)^2,\label{eq:mediumaver}
\eeq
which can be written as
\beq
\langle|{\cal M}_2^{a_2a_1a}|^2\rangle=4g^2C_F\sum_{n_1,n_2}\left(\frac{\sigma^{\rm el}}{A_\perp}\right)^2\int d\q_1\left(\frac{1}{\sigma^{\rm el}}\frac{d\sigma^{\rm el}}{d\q_1}\right)\int d\q_2\left(\frac{1}{\sigma^{\rm el}}\frac{d\sigma^{\rm el}}{d\q_2}\right)\,|\J_1|^2\,(2p_+)^2\, .
\eeq
Here, $\sigma^{\rm el}$ denotes the quark elastic cross section.
One has still to perform an average over the longitudinal position of the scattering centers. Notice that in the above we assumed $x_{n_2}^+>x_{n_1}^+$ which implies
\beq
\sum_{n_1,n_2}\left(\frac{1}{L^+}\right)^2\int_0^{L^+} dx_{n_1}^+\int_{x_{n_1}^+}^{L^+} dx_{n_2}^+=\left(\frac{N}{L^+}\right)^2\frac{1}{2}(L^+)^2=\frac{N^2}{2}\, .
\eeq
Recasting this expression in the form 
\beq
\left(\frac{\sigma^{\rm el}}{A_\perp}\right)^2\frac{N^2}{2}=\left(\frac{\sigma^{\rm el}N}{A_\perp L^+}\right)^2\frac{(L^+)^2}{2}=\left(\frac{L^+}{\lambda_{q}^+}\right)^2\frac{1}{2}\, ,
\eeq
one finds 
\beq
\langle|{\cal M}_{a_2a_1a}|^2\rangle=4g^2C_F\left(\frac{L^+}{\lambda_{q}^+}\right)^2\,|J_{-\infty}|^2\,\left\langle\left\langle\frac{1}{2}|\J_1|^2\right\rangle\right\rangle_{\q_1,\q_2},
\eeq
which coincides with the expression quoted in the text.
Finally, the longitudinal average of terms containing interference factors (which appear in other channels) can be performed exploiting
\beq
\sum_{n_1,n_2}\left(\frac{1}{L^+}\right)^2\int_0^{L^+} dx_{n_1}^+\int_{x_{n_1}^+}^{L^+} dx_{n_2}^+\cos[\Omega(x_{n_2}^+-x_{n_1}^+)]=N^2\frac{1-\cos[\Omega L^+]}{(\Omega L^+)^2}.
\eeq

The medium average Eq~(\ref{eq:colortrace}) is the same as employed in standard color inclusive calculations. It traces independently over the color 
of the scattering centers $n_1$ and $n_2$. We emphasize that in the large-$N_c$ limit this procedure does not affect the decomposition of the radiation spectrum
into distinct color channels. Here, we consider for simplicity the case in which the scattering of the projectile occurs on a medium particle in the (anti-)fundamental representation.
After averaging over all the incoming colors, one has, according to Eq.~(\ref{eq:mediumaver}), the overall color factor
\beq
C_F\frac{C_FT_{F}}{d_{F}}\frac{C_FT_{F}}{d_{F}}=C_F\frac{C_F}{2N_c}\frac{C_F}{2N_c}\sim\frac{N_c}{32}.\label{eq:colorfactor}
\eeq
This result can also be obtained graphically from Fig.~\ref{fig:N2flow} with the simple rule that each closed loop produces a factor $N_c$, and each gluon exchange, due to the Fierz identity
\beq
t^a_{ij}t^a_{kl}=\frac{1}{2}\delta_{il}\delta_{jk}-\frac{1}{2N_c}\delta_{ij}\delta_{kl}\sim\frac{1}{2}\delta_{il}\delta_{jk}\, ,
\eeq
gives rise in the large-$N_c$ limit to a factor $1/2$. One gets then for Fig.~\ref{fig:N2flow}, after averaging over the initial colors:
\beq
\frac{1}{N_c^3}N_c^4\left(\frac{1}{2}\right)^5=\frac{N_c}{32},
\eeq
in agreement with Eq.~(\ref{eq:colorfactor}).
\subsection{Virtual contribution}
Let us now address the evaluation of the $2{\rm Re}\langle{\cal M}_3^{\rm virt}{\cal M}_1^*\rangle$ contribution. The amplitude ${\cal M}_3$ interferes either with the initial-state radiation ($a_ia$ channel)
\beq
i{\cal M}_{1}^{\rm in}\equiv i{\cal M}_{1}^{a_ia} =(ig)\sum_i(t^{a_i}t^a)\!\!\int\!\!\!\frac{d\q'}{(2\pi)^2}e^{-i\q'\cdot\x_i}\, 2\,\eeg\!\cdot\!\J(\kg,q')\,e^{i\wob x_i^+}\,(2p_+)\,{\cal A}(\q')\,T_{(i)}^{a_i}\, ,
\eeq
or with the final-state radiation ($aa_1$ channel)
\beq
i{\cal M}_{1}^{\rm fin}\equiv i{\cal M}_{1}^{aa_i}=-(ig)\sum_i(t^a t^{a_i})\!\!\int\!\!\!\frac{d\q'}{(2\pi)^2}e^{-i\q'\cdot\x_i}\,2\,\eeg\!\cdot\!\J(\kg,q')\,e^{i\wob x_i^+}\,(2p_+)\,{\cal A}(\q')\,T_{(i)}^{a_i}\, .
\eeq
The various terms contributing to ${\cal M}_3^{\rm virt}$ can be obtained from the $N\!=\!3$ diagrams listed in~\cite{Gyulassy:1999zd}. While each individual diagram can 
produce up to four different contributions ($x_1\!=\!x_2$ or $x_2\!=\!x_3$, interference with ${\cal M}_{1}^{\rm in}$ and ${\cal M}_{1}^{\rm fin}$), only some of them yield a 
leading contribution in the large-$N_c$ limit. Their expression can be directly read from~\cite{Gyulassy:1999zd} after taking properly the contact limit of two of the three scattering centers. 
To fix the exact overall factors it is sufficient to consider one case, e.g. the term with the gluon radiated before the interaction with the first scattering center and with a double interaction at the second one
(see Fig.~\ref{fig:N3contact}). This corresponds to the first of the 15 diagrams shown in Appendix C of Ref.~\cite{Gyulassy:1999zd} (hence we label it as `$I$'). After setting $x_2\!\equiv\!x_3$ and considering the projectile arriving on-shell, one has
\begin{multline}
i{\cal M}_{a'_ja_ja_ia}^{\rm virt\,(I)}=\sum_{i<j}g(t^{a'_j}t^{a_j}t^{a_i}t^{a})(T_{(j)}^{a'_j}T_{(j)}^{a_j})T_{(i)}^{a_i}\int\!\!\frac{d\q_1}{(2\pi)^2}\int\!\!\frac{d\q_2}{(2\pi)^2}\int\!\!\frac{d\q_3}{(2\pi)^2}\int\!\!\frac{dq_{23}^-}{2\pi}\int\!\!\frac{dq_{3}^-}{2\pi} \\
\times e^{-i\q_1\cdot\x_i}e^{-i(\q_2+\q_3)\cdot\x_j} e^{i(q_1+q_2+q_3)^-x_i^+}e^{i(q_2+q_3)^-(x_j-x_i)^+}(i)[(2p_f-q_3)^+{\cal A}(\q_3)]\\
\times\frac{i}{(p_f-q_3)^2-M^2+i\eta}(i)[(2p_f-2q_3-q_2)^+{\cal A}(\q_2)]\frac{i}{(p_f-q_2-q_3)^2-M^2+i\eta}\\
\times(i)[(2p_f-2q_3-2q_2-q_1)^+{\cal A}(\q_1)]\frac{i}{(p_i-\kg)^2-M^2+i\eta}(i)[(2p_i\!-\!\kg)\cdot\eg],
\end{multline}   
where $q_{23}^-\!\equiv\!q_2^-\!+\!q_3^-$. Furthermore, from the on-shell condition and momentum conservation one has
\beq
(q_1\!+\!q_2\!+\!q_3)^-=\frac{\kkg^2}{2\xg p^+}+\frac{(\Q\!-\!\kkg)^2}{2(1\!-\!\xg)p^+}-\frac{M^2}{2p^+}\approx\!\wob\, .\quad\quad(\Q\equiv\q_1+\q_2+\q_3)
\eeq
This fixes $q_1^-$.
The integration over $q_{23}^-$ can be evaluated picking the pole in the complex upper half-plane of the propagator
\beq
\frac{i}{(p_f-q_2-q_3)^2-M^2+i\eta}=\frac{(-i)}{2(1-\xg)p^+\left[q_{23}^--\left(\frac{(\Q-\kkg)^2-(\q_1-\kkg)^2}{2(1-\xg)p^+}+i\eta\right)\right]}\, ,
\eeq
which in the high-energy limit lies at $q_{23}^-\approx i\eta$. For what concerns the integration over $q_3^-$, in the contact limit, no phase factor is present and one no longer closes the contour in the complex plane. However one can rely on the following representation of the Dirac delta (no contribution to the integration arises from the imaginary part):
\begin{multline}
\frac{i}{(p_f-q_3)^2-M^2+i\eta}=\frac{(-i)}{2(1-\xg)p^+\left[q_{3}^--\left(\frac{(\Q-\kkg)^2-(\q_1+\q_2-\kkg)^2}{2(1-\xg)p^+}+i\eta\right)\right]}\\
=\frac{\pi}{2(1-\xg)p^+}\,\frac{1}{\pi}\frac{\eta}{\left[q_{3}^--\frac{(\Q-\kkg)^2-(\q_1+\q_2-\kkg)^2}{2(1-\xg)p^+}\right]^2+\eta^2}\approx\frac{\pi}{2(1-\xg)p^+}\delta(q_3^-).
\end{multline}
We thus obtain in the soft limit
\begin{multline}
i{\cal M}_{a'_ja_ja_ia}^{\rm virt\,(I)}\approx(-i)\frac{g}{2}\sum_{i<j}g(t^{a'_j}t^{a_j}t^{a_i}t^{a})(T_{(j)}^{a'_j}T_{(j)}^{a_j})T_{(i)}^{a_i}\int\!\!\frac{d\q_1}{(2\pi)^2}\int\!\!\frac{d\q_2}{(2\pi)^2}\int\!\!\frac{d\q_3}{(2\pi)^2}\\
\times e^{-i\q_1\cdot\x_i}e^{-i(\q_2+\q_3)\cdot\x_j}e^{i\wob\x_i^+}{\cal A}(\q_3){\cal A}(\q_2)\,(2p_+)\,{\cal A}(\q_1)\left(2\eeg\!\cdot\!\frac{\kkg}{\kkg^2+\xg^2M^2}\right).
\end{multline}
We now evaluate the interference with the $N\!=\!1$ amplitude, taking as usual the proper medium average (over colors, transverse and longitudinal location of the scattering centers). The contribution from the final-state radiation channel is vanishing, 
\beq
2{\rm Re}\langle{\cal M}_{a'_ja_ja_ia}^{\rm virt\,(I)}({\cal M}_1^{\rm fin})^*\rangle=0\, .
\eeq
For the interference with the initial state radiation one gets
\beq
2{\rm Re}\langle{\cal M}_{a'_ja_ja_ia}^{\rm virt\,(I)}({\cal M}_1^{\rm in})^*\rangle=-4g^2C_F\left(\frac{L^+}{\lambda_{q}^+}\right)^2\,|J_{-\infty}|^2\,\frac{1}{2}\left\langle\left\langle\frac{\kkg}{\kkg^2+\xg^2M^2}\!\cdot\!\J(\kg,q_1)\right\rangle\right\rangle_{\q_1,\q_2}\, .
\eeq
The other contributions can be evaluated analogously.

\providecommand{\href}[2]{#2}\begingroup\raggedright\endgroup


\end{document}